\documentclass[11pt,twoside,letterpaper]{article} 
\usepackage{times,fancyhdr}
\usepackage[dvips]{graphicx}
\usepackage{color}
\usepackage{amsfonts}
\usepackage{url}
\usepackage{subfigure}

\makeatletter 
\def\cleardoublepage{\clearpage\if@twoside \ifodd\c@page\else%
    \hbox{}%
    \thispagestyle{empty}%
    \newpage%
    \if@twocolumn\hbox{}\newpage\fi\fi\fi} 
\makeatother

\renewcommand{\thesection}{\arabic{section}.}
\renewcommand{\thesubsection}{\thesection\arabic{subsection}.}

\def\figurename{Figure}
\makeatletter
\renewcommand{\fnum@figure}[1]{\figurename~\thefigure.}
\makeatother

\def\tablename{Table}
\makeatletter
\renewcommand{\fnum@table}[1]{\tablename~\thetable.}
\makeatother

\begin{document}
\title{
{\begin{flushleft}
\vskip 0.45in
{\normalsize\bfseries\textit{Chapter~1}}
\end{flushleft}
\vskip 0.45in
\bfseries\scshape Wavelets for Density-Functional Theory and Post-Density-Functional-Theory Calculations}}
\author{\bfseries\itshape Bhaarathi Natarajan$^{a,b}$\thanks{E-mail address: Bhaarathi.Natarajan@UJF-Grenoble.FR}\\
\bfseries\itshape Mark E.\ Casida$^a$\thanks{E-mail address: Mark.Casida@UJF-Grenoble.FR}\\
\bfseries\itshape Luigi Genovese$^{b}$\thanks{E-mail address: Luigi.Genovese@cea.FR}\\
\bfseries\itshape Thierry Deutsch$^{b}$\thanks{E-mail address: Thierry.Deutsch@CEA.FR}\\
$^a$ Laboratoire de Chimie Th\'eorique,\\
D\'epartement de Chimie Mol\'ecularie (DCM, UMR CNRS/UJF 5250),\\
Institut de Chimie Mol\'eculaire de Grenoble (ICMG, FR2607),\\
Universit\'e Joseph Fourier (Grenoble I),\\
301 rue de la Chimie, BP 53,\\
F-38041 Grenoble Cedex 9, France\\
$^b$ UMR-E CEA/UJF-Grenoble 1,\\
INAC, Grenoble, F-38054, France\\}
\date{\today} 
\maketitle
\thispagestyle{empty}
\setcounter{page}{1}
\thispagestyle{fancy}
\fancyhead{}
\fancyhead[L]{In: Theoretical and Computational Methods in Modern\\ Density Functional Theory \\ 
Editor: A.K.\ Roy, pp. {\thepage-\pageref{lastpage-01}}} 
\fancyhead[R]{ISBN 0000000000  \\
\copyright~2011 Nova Science Publishers, Inc.}
\fancyfoot{}
\renewcommand{\headrulewidth}{0pt}

\vspace{2in}

\vspace{.08in} \noindent \textbf{Keywords:} Wavelets, density-functional theory, time-dependent density-functional theory,
linear-response time-dependent density-functional theory, orbital energies, electronic excitation energies.

\pagestyle{fancy}
\fancyhead{}
\fancyhead[EC]{Natarajan, Genovese, Casida, and Deutsch}
\fancyhead[EL,OR]{\thepage}
\fancyhead[OC]{Wavelets for DFT and Post-DFT Calculations}
\fancyfoot{}
\renewcommand\headrulewidth{0.5pt} 
\begin{abstract}
We give a fairly comprehensive review of wavelets
and of their application to density-functional theory (DFT) and to our recent
application of a wavelet-based version of linear-response time-dependent DFT (LR-TD-DFT).
Our intended audience is quantum chemists and theoretical solid-state and chemical
physicists.  Wavelets are a Fourier-transform-like approach which developed primarily
in the latter half of the last century and which was rapidly adapted by 
engineers in the 1990s because of its advantages compared to standard
Fourier transform techniques for multiresolution problems with complicated
boundary conditions.  High performance computing wavelet codes now also exist
for DFT applications in quantum chemistry and solid-state physics, notably the
{\sc BigDFT} code described in this chapter.  After briefly describing the basic
equations of DFT and LR-TD-DFT, we discuss how they are solved in {\sc BigDFT}
and present new results on the small test molecule carbon monoxide to show how
{\sc BigDFT} results compare against those obtained with the quantum chemistry
gaussian-type orbital (GTO) based code {\sc deMon2k}.  In general, the two programs
give essentially the same orbital energies, but the wavelet basis of {\sc BigDFT}
converges to the basis set limit much more rapidly than does the GTO basis set of
{\sc deMon2k}.  Wavelet-based LR-TD-DFT is still in its infancy, but our calculations
confirm the feasibility of implementing LR-TD-DFT in a wavelet-based code. 
\end{abstract}
\tableofcontents
\section{Introduction}
\label{sec:intro}

The broad meaning of ``adaptivity'' is the capacity to make something work better by alternation, 
modification, or remodeling.
Concepts of adaptivity have found widespread use in quantum chemistry,
ranging from the construction of Gaussian-type orbital (GTO) basis sets,   
see e.g., the development of correlation consistent bases \cite{D89,KDH92,WD93}, 
to linear scaling methods in density functional theory (DFT)  
\cite{LNV93,D93,GC94,K96,MS97,OH97}, 
selective configuration interaction (CI)  methods \cite{HMR73,EDM83} and   
local correlation methods based on many-body perturbation theory  or
coupled cluster (CC)  theory \cite{SHW99, SA99}.  
This chapter is about a specific adaptive tool, namely wavelets as an adaptive basis set 
for DFT calculations which can be automatically placed when and where needed to handle multiresolution 
problems with difficult boundary conditions.  

Let us take a moment to contrast the wavelet concept of adaptivity with other types of adaptivity.
In other contexts, the adaptive procedure is typically based on a combination of physical insights together
with empirical evidence from numerical simulations. A rigorous mathematical justification
is usually missing. This may not be surprising: Familiar concepts lose a lot of their original
power if one tries to put them in a rigorous mathematical framework. 
Therefore, we will not shoulder the monumental and perhaps questionable task
of providing a rigorous mathematical analysis of all the adaptive approaches used
nowadays throughout quantum chemistry.  Instead we will concentrate on the mathematical analysis of
a particular electronic structure method which lends itself to a rigorous mathematical analysis and
application of adaptivity.
In contrast with other adaptive methods, multiresolution analysis (MRA) with wavelets 
can be regarded as an additive subspace correction and their wavelet representations have a naturally 
built-in adaptivity which comes through their ability to express directly
and separate components of the desirable functions living on different scales.

This combined with the fact
that many operators and their inverses have nearly sparse representations in wavelet coordinates may eventually
lead to very efficient schemes that rely on the following principle: Keep the computational work proportional to the
number of significant coefficients in the wavelet expansions of the searched solution. As there are a lot of different
wavelet bases with different properties (length of support, number of vanishing moments, symmetry, etc.) in each
concrete case we can choose the basis that is most appropriate for the intrinsic complexity of the sought-after solution.
This fact makes the wavelet-based schemes a very sophisticated and powerful tool for compact representations of
rather complicated functions. 
The expected success of wavelet transforms for solving electronic structure
problems in quantum mechanics are due to three important properties: (a) the ability to choose a basis set providing
good resolution where it is needed, in those cases where the potential energy varies rapidly in some regions of space,
and less in others; (b) economical matrix calculations due to their sparse and banded nature; and (c) the ability to
use orthonormal wavelets, thus simplifying the eigenvalue problem.
Of course, this might lead to adaptive methods which are fully competitive
from a practical point of view, for example, working with a systematic basis
instead of GTO bases requires from the onset larger
basis sets and the benefit of systematic improvement might be a distant
prospect.  However, we have the more realistic prospect that our rigorous
analysis provides new and hopefully enlightening perspectives on standard 
adaptive methods, which we reckon cannot be obtained in another way.

On the otherhand advances in computational technology opened up new opportunities 
in quantum mechanical calculation of various electronic structures, 
like molecules, crystals, surfaces, mesoscopic systems, etc. 
The calculations can only be carried out either for very limited systems 
or with restricted models, because of their great demand of 
computational and data storage resources.
Independent particle approximations, like the Hartree-Fock based \cite{H57,F30,S30,R51}
algorithms with single determinant wave functions, 
leave out the electron correlation and need operation and 
storage capacity of order $N^4$, if $N$ is the
total number of electrons in the system. If inclusion of the electron correlation
is necessary, CI or CC
methods can be applied, with very high demand of computational resources
(${\cal{O}}(N^6)$ to ${\cal{O}}(N!)$). 
An alternative way is to use MBPT. 
The second order perturbation calculations can be carried out 
within quite reasonable time and resource limits, 
but the results are usually unsatisfactory, they just show the tendencies, 
while the 4th order MBPT needs ${\cal{O}}(N^7)$ to ${\cal{O}}(N^8)$ operations.
All these algorithms use the $N$-electron wave function as a basic quantity.

Another branch of methods use electron density as the primary entity.
Pioneers of this trend, like Thomas \cite{T26}, Fermi \cite{F26,F26a}, Frenkel 
\cite{F28} and Sommerfeld \cite{S28} developed the statistical theory of 
atoms and the local density approximation (LDA).   
The space around the nuclei is separated into small regions,
where the atomic potential is approximated as a constant, 
and the electrons are modeled as a free electron gas of 
Fermi-Dirac statistics \cite{D26,F27,F28}. Dirac included electron correlation 
\cite{D30}, which improved the results.
After the Hohenberg--Kohn theorems had appeared \cite{HK64}, and Kohn and
Sham had offered a practically applicable method \cite{KS65} 
based on their work, many scientists were motivated to work on the theory, 
and DFT developed into one of the most powerful electronic structure methods.

Despite the success of density functional theory, it has some drawbacks.
The exact formula of the exchange-correlation potential is not known, 
thus chemical intuition and measured data are necessary 
in order to approximate it, and the kinetic energy functional 
is hard to calculate. Powerful approximating formulas are available 
(see, e.g. \cite{DG90}), like the Thomas--Fermi functional based gradient 
and generalized gradient expansions, where the energy functionals 
are expressed as a power series of the gradient of the density 
(the first such suggestion was \cite{HVO69}.)

Considering the historical development of sophisticated 
$N$-electron methods, a typical trend can be observed. 
Starting with a very simple model, new details are introduced 
in order to improve the results. This scheme is
followed in the linear muffin tin orbital method (LMTO) \cite{A75}  
where the interatomic regions 
is replaced by the spherical orbital of an atomic potential 
around the nuclei. 
Similarly, the linearized augmented plane wave method (APW) \cite{WKWF81}  
and the plane wave pseudopotential approach \cite{PTAAJ92} 
describe the details of the crystal potential differently 
in different spatial regions. Although they are rather successful, 
for applying any of these models, chemical intuition is needed, 
free parameters, like the radius of the bordering sphere between
the two types of potentials, and the boundary conditions have to be set.
A systematic method, which can handle the different behaviors 
of the electron structures at different spatial domains, 
or either at different length scale \cite{PV97}, is 
the longterm requirement of any  physical chemists.

Multiresolution or wavelet analysis, this rapidly developing
branch of the applied mathematics, is exactly the tool for 
statisfy all the need of any chemical physicists/physcial chemists.
From mathematical point of view wavelet analysis  is a
theory of a special kind of Hilbert space basis sets. 
Basis sets are commonly used in all electron structure calculations, 
as the wave function is usually expanded as linear combination 
of some kind of basis functions. 
Thus the operator eigenvalue problem 
is reduced to an algebraic matrix eigenvector problem. 
The resulting algebraic equations are easier to solve, well known
algorithms and subroutine libraries are available, however, 
the difficulty of choosing the proper basis set arises. 
If linear combination of atomic orbitals (LCAO)  is used,  
the atomic basis functions are Slater or Gaussian-type of
functions \cite{PKC85,PB70}, the selection of atomic orbitals needs chemical 
intuition, which is a result of long time's experience, and can not 
be algorithmized. Both basis sets are non-orthogonal, 
and lack the explicit convergence properties \cite{WC96}.
Moreover, calculation of operator matrix elements with Slater-type
orbitals is complicated, their integrals have to be treated numerically. 
Although integrals of Gaussian functions are analytically known, 
the Gaussisn-type basis does not reflect the nuclear cusp condition of Kato
\cite{K57}, which reflects on 
singularities of the $N$-electron wave function in the presence of 
Coulomb-like potentials. Since then it turned out that for high precision 
numerical calculations it is essential to satisfy these
requirements. However, while the nuclear cusp condition is relatively 
easy to fulfill by Slater-type orbitals (STO),   
the electron-electron cusp is extremely hard to represent.
In general, GTO-based/STO-based DFT codes gives reliable results with a
relatively small number of basis functions, making them optimal for large
scale computations where high accuracy is less crucial.  On the other hand there
is no consistent way to extend these basis sets and thereby converge the 
results with respect to the size of the basis.

The second type of basis set covers the system-independent functions such as
plane waves \cite{PTAAJ92} or wavelets \cite{ALE97}.  The main advantable of these
basis sets, is that their size can be systematically increased until the 
result of the calculation has converged, and are generally considered to 
be more accurate than the former type.  The number of basis function 
required to obtain convergence is normally {\em so large} that direct
solution of the matrix eigenvalue problem within the entire basis space is 
{\em not} 
possible.  Instead one has to use iterative methods to determine the lowest 
(occpied) part of the spectrum \cite{PTAAJ92}.
In solid state physics, where more or less periodic systems are studied,
choosing plane wave basis sets is rather usual. 
These basis functions are system independent and easily computable, 
but the results are not always convincing and the number of necessary 
basis functions is almost untreatable.
(Theoretically, plane waves could also be used for describing molecules,
since the two-electron integrals and the expectation values 
are connected to the Fourier transform, thus they are easily computable, 
and this could balance the large number of necessary basis functions.) 
The reason, why so many plane waves are needed is that the 
wave functions around the nuclei need very high frequency terms, 
i.e. high resolution level, for reproducing the nuclear cusps. 
In the framework of Fourier analysis, the whole space has
to be expanded at the same resolution, despite that in most of the 
space low frequency terms would be sufficient.

Fully-numerical ``basis-set free'' Hartree-Fock (HF) calculations of atoms have been known since the 1960s
(Vol. 1, pp. 322-326 and Vol. 2, pp. 15-30 of Ref.~\cite{S60} and Refs.~\cite{F63,F72,F73,M73})
and have proven helpful in constructing efficient finite basis sets for molecular
calculations.  
In the late 1980s, Axel Becke used a fully-numerical density-functional theory (DFT) program for diatomics to show that
many of the problems of DFT calculations at that time were due not to the functionals used, but rather numerical
artifacts of the DFT programs of the 1970s ~\cite{HFK88}.)
Since that time, fully-numerical DFT codes have been implemented
for polyatomic molecules using the finite element method (FEM), with {\sc Parsec} from the 
chemists point of view or {\sc OCTOPUS} from the view of physicsts being a notable example.

{\sc BigDFT} the pseudo potential code for bigger systems
based as it is on traditional Hohenberg-Kohn-Sham DFT \cite{HK64,KS65},
could only calculate ground-state properties with an eye to order-N DFT. As a step to increase
the feasibility of the code we formulated the wavelet-based linear-response time-dependent
density-functional theory (TD-DFT) and here we support our first implementation for calculating 
electronic excitation spectra \cite{NGC+11}.
Electronic excitation spectra can be calculated from TD-DFT \cite{RG84} using time-dependent linear response (LR) theory 
\cite{PGG96,C95}.  Casida formulated  LR-TD-DFT (often just refered to as TD-DFT) so as to resemble the linear-response
time-dependent HF equations already familiar to quantum chemists \cite{C95}.  That method was then rapidly implemented
in a large number of electronic structure codes in quantum chemistry, beginning with the {\sc deMon} family of programs
\cite{JCS96} and the {\sc TurboMol} program \cite{BA96}.  Among the programs that implemented ``Casida's equations''
early on was the FEM DFT program {\sc PARSEC} \cite{CKSL08} and also be found in the FEM DFT program {\sc Octopus} \cite{CMA+06}.
  See Ref.~\cite{LHP09} for a recent FEM implementation
of TD-DFT. Since a wavelet-based program offers certain advantages
over these other FEM DFT programs, it was deemed important to also implement LR-TD-DFT in {\sc BigDFT}.

In the next section we give a detailed description of the idea behind the multiresolution analysis and wavelets, with a historical note.
Sec.~\ref{sec:dft} and Sec.~\ref{sec:tddft}, briefly presenting the theoretical introduction to DFT and
TD-DFT, and Sec.~\ref{sec:krylov},
talks about the well-known Krylov space methods for solving eigenvalue
equations involved in our implementation. 
Sec.~\ref{sec:bigdft-dft} and Sec.~\ref{sec:bigdft-tddft}, 
gives the numerical implementation
of DFT and how we have implemented TD-DFT from the aspects of theoretical and algorithmic
point of view on wavelets based  pseudopotential electronic structure code {\sc BigDFT},
and in Section \ref{sec:results} we give the results of detailed
comparisons between TD-DFT excitation spectra calculated with {\sc BigDFT} and with the implementation of Casida's
equations in the GTO-based program {\sc deMon2k}. The conclusion were drawn
for future applications in the field of chemistry and some of the other
problems are reviewed to draw chemists' greater attention to wavelets 
and to gain more benifits from using wavelet technique.

\section{Wavelet Theory}
\label{sec:wavelets}
The mathematics of wavelets is a fairly new technique,
it can generally be used where one traditionally uses
Fourier techniques.  They incorporate the feature of having
multiple scales, so very different resolutions
can be used in different parts of space in a mathematically rigorous
manner.  This matches many systems in nature well, for example
molecule where the atomic orbitals are very detailed close to the
cores, while they only vary slowly between them.
Wavelet analysis can quite generally be viewed as a
local Fourier analysis. From the wavelet expansion, or wavelet spectrum,
of a function, $f$, it can be inferred not only how fast $f$ varies,
i.e. which frequencies it contains, but also where in space a given frequency
is located. This property has important applications
in both data compression, signal/image processing and noise reduction
\cite{N98}. Wavelet methods are also employed for solving partial differential
equations \cite{DCD01,J03}, and in relation to electronic structure methods
a complete DFT program based on interpolating wavelets has been developed \cite{BigDFT}.

\subsection{The story of wavelets}
Most historical versions of wavelet theory however, despite their source's
perspective, begin with Joseph Fourier.
In 1807, a French mathematician, Joseph Fourier,
discovered that all periodic functions could be expressed
as a weighted sum of basic trigonometric functions.
His ideas faced much criticism from Lagrange, Legendre
and Laplace for lack of mathematical rigor and
generality, and his papers were denied publication. It
took Fourier over 15 years to convince them and publish
his results. Over the next 150 years his ideas
were expanded and generalized for non-periodic
functions and discrete time sequences. The fast Fourier
transform algorithm, devised by Cooley and
Tukey in 1965 placed the crown on Fourier transform,
making it the king of all transforms. Since then
Fourier transforms have been the most widely used,
and often misused, mathematical tool in not only
electrical engineering, but in many disciplines requiring
function analysis.
This crown however, was about to change hands. Following
a remarkably similar history of development, the
wavelet transform is rapidly gaining popularity and recognition.

The first mention of wavelets was in a 1909 dissertation by Hungarian
mathematician Alfred Haar. Haar's work was not necessarily about wavelets,
as ``wavelets" would not appear in their current form until the late 1980s.
Specifically, Haar focused on orthogonal function systems, and proposed an
orthogonal basis, now known as the Haar wavelet basis, in which functions
were to be transformed by two basis functions. One basis function is
constant on a fixed interval, and is known as the scaling function.
The other basis function is a step function that contains exactly one
zero--crossing (vanishing moment) over a fixed interval (more on this later).

The next major contribution to wavelet theory was from a 1930s French scientist
Paul Pierre L\'evy. More correctly, L\'evy's contribution was less of
a contribution and more of a validation. While studying the ins and outs
of Brownian motion in the years following Haar's publication, L\'evy
discovered that a scale--varied Haar basis produced a more accurate
representation of Brownian motion than did the Fourier basis. L\'evy, being
more of a physicist than mathematician, moved on to make large
contributions to our understanding of stochastic processes.

Contributions to wavelet theory between the 1930s and 1970s were slight.
Most importantly, the windowed Fourier transform was developed, with the
largest contribution being made by another Hungarian
named Dennis Gabor. The next major advancement in wavelet theory is
considered to be that of Jean Morlet in the late 1970s.

Morlet, a French geophysicist working with windowed Fourier transforms,
discovered that fixing frequency and stretching or compressing (scaling)
the time window was a more useful approach than varying frequency and
fixing scale. Furthermore, these windows were all generated
by dilation or compression of a prototype Gaussian.  These window
functions had compact support both in time and in frequency (since the
Fourier transform of a Gaussian is also a Gaussian.) Due to
the small and oscillatory nature of these window functions, Morlet
named his functions as ``wavelets of constant shape". In 1981, Morlet worked with Croatian--French physicist
Alex Grossman on the idea that a function could be transformed by a
wavelet basis and transformed back without loss of information, thereby
outlining the wavelet transformation. It is of note that Morlet initially
developed his ideas with nothing more than a handheld calculator.

In 1986, St\'ephane Mallat noticed a publication by Yves Meyer that
built on the concepts of Morlet and Grossman. Mallat sought Meyer's
consult, and the result of said consult was Mallat's publication of
multiresolution analysis. Mallat's MRA connected wavelet
transformations with the field of digital signal processing.
Specifically, Mallat developed the wavelet transformation as a
multiresolution approximation produced by a pair of digital filters.
The scaling and wavelet functions that constitute a wavelet basis are
represented by a pair of finite impulse response filters, and the
wavelet transformation is computed as the convolution of these filters
with the input function. The importance of Mallat's contribution cannot be
overstated. Without the fast computational means of wavelet transformation
provided by the MRA, wavelets, undoubtedly, would not be the effective and
widely used signal processing tools that they are today.

In 1988, a student of Alex Grossman, named Ingrid Daubechies, combined the
ideas of Morlet, Grossman, Mallat, and Meyer by developing the first family
of wavelets as they are known today. Named the Daubechies wavelets,
the family consists of 8 separate wavelet and scaling functions (more on this later).
With the development of pair Daubechies wavelet and scaling functions is
orthogonal, continuous, regular, and compactly supported, the foundations
of the modern wavelet theory were laid.  The last ten years mostly
witnessed a search for other wavelets with different properties and
modifications of the MRA algorithms.  In 1992, Albert Cohen, Jean Feauveau
and Daubechies constructed the compactly supported biorthogonal
wavelets, which are preferred by many researchers over the
orthonormal basis functions, whereas R. Coifman, Meyer and Victor
Wickerhauser developed wavelet packers, a natural extension of MRA.
\subsection{Multiresolution analysis }
A suitable gateway to the theory of wavelets is through 
the idea of MRA. 
A detailed description of MRAs can be found in Keinert \cite{K03}, 
from which a brief summary of the key issues are given in the following.

A multiresolution analysis is an infinite nested sequence of 
subspaces $L^2({\mathbb{R}})$
\begin{equation}
V_j^{0} \subset V_j^1 \subset ...\subset V_j^n \subset ...
\label{eq:mra.1}
\end{equation}
with the following properties
\begin{itemize}
\item {$ V_j^{\infty} $ is dense in $L^2 $ }
\item{$f(x) \in V_j^n \Longleftrightarrow f(2x) \in V_j^{n+1} \mbox 0\leq n \leq \infty$}
\item{$f(x) \in V_j^n \Longleftrightarrow f(x-2^{-n} l) \in V_j^{n} \mbox 0\leq l \leq (2^n-1)$}
\item{ There exists a function vector $\varphi$ of length $j+1$ in $L^2$ such that 
$$\{\varphi_j(x) :\mbox 0\leq k \leq j \} $$ forms a basis for $V_j^0.$}
\end{itemize}

This means that if we can construct a basis of $V_j^0$, which consists of only
$j+1$ functions, we can construct a basis of any space $V_j^n$, by simple
compression (by a factor of $2^n$), and translations (to all grid points
at scale $n$), of the original $j+1$ functions, and by increasing the
scale $n$, we are approaching a complete basis of $L^2$.  Since
$V_j^n \subset V_j^{n+1}$ the basis functions of $V_j^n$ can be expanded
in the basis of $V_j^{n+1}$
\begin{equation}
\varphi_l^n (x) \begin{array}{c} \mbox{def} \\ = \end{array} 2^{n/2} \varphi(2^n x -l) = \sum_l h^{(l)} \varphi_l^{n+1} (x) \, .
\label{eq:mra.2}
\end{equation}
where $h^{(l)}$s are the so-called filter matrix that describes the transformation
between different spaces $V_j^n$.

The MRA is called orthogonal if
\begin{equation}
\langle \varphi_0^n(x), \varphi_l^n(x)\rangle = \delta_{0l} I_{j+1} \, ,
\label{eq:mra.3}
\end{equation}
where $I_{j+1}$ is the $(j+1)\times (j+1)$ unit matrix, and $j+1$ is the 
length of the function vector.  The orthogonality condition means that 
the functions are orthogonal both within one function vector and through
all possible translations on one scale, but not through the
different scales.

Complementary to the nested sequence of subspaces $V_j^n$, we can define
another series of spaces $W_j^n$ that complements $V_j^n$ in $V_j^{n+1}$
\begin{equation}
V_j^{n+1} = V_j^n \oplus W_j^n
\label{eq:mra.4}
\end{equation}
where there exists another function vector $\phi$ of length $j+1$ that, 
with all its translations on scale $n$ forms a basis for $W_j^n$.
Analogously to Eq.~(\ref{eq:mra.2}) the function vector can be
expanded in the basis of $V_j^{n+1}$
\begin{equation}
\phi_l^n (x) \begin{array}{c} \mbox{def} \\ = \end{array}  2^{n/2} \phi(2^n x -l) = \sum_l g^{(l)} \phi_l^{n+1} (x) \, .
\label{eq:mra.5}
\end{equation}
with filter matrices $g^{(l)}$. The functions $\phi$ also fulfill the same
orthogonality condition as Eq.~(\ref{eq:mra.3}), and if we combine Eq.~(\ref{eq:mra.1}) and Eq.~(\ref{eq:mra.4}) we see that they must be 
orthogonal with respect to different scales.  Using Eq.~(\ref{eq:mra.4})
recursively we obtain
\begin{equation}
V_j^n = V_j^0 \oplus W_j^0 \oplus W_j^1 \oplus ...\oplus W_j^{n-1} \, .
\label{eq:mra.6}
\end{equation}
which will prove to be an important relation.
\subsection{Wavelets}
There are many ways to choose the basis functions $\varphi$ and $\phi$
(which define the spanned spaces $V_j^n$ and $W_j^n$), and
there have been constructed functions with a variety of properties, and
we should choose the wavelet family that best suits the needs of the
problem we are trying to solve.  (Wavelets are often denoted by $\psi$ in the literature
but the choice has been made here to denote them by $\phi$ so as to avoid confusion with
the Kohn-Sham orbitals.) Otherwise, we could start from scratch
and construct the new family, one that is custum-made for the problem at
hand.  Of course, this is not a trivial task, and it might prove more
efficient to use an existing family, even though its properties are not
right on cue.

There is a one-to-one correspondence between the basis functions $\varphi$
and $\phi$, and the filter matrices $h^{(l)}$ and $g^{(l)}$ used in the
two-scale relation equations Eq.~(\ref{eq:mra.2}) and Eq.~(\ref{eq:mra.5}),
and most well-known wavelet families are defined only through their
filter coefficients. 

In the following we are taking a different, more intuitive approach, for
defining the {\em scaling space} $V_j^n$ as the space of 
piecewise polynomial functions
\begin{eqnarray}
V_j^n & \begin{array}{c} \mbox{def} \\ = \end{array} & \{f: \mbox{} \mbox{ all polynomials of degree} \leq j  \, \nonumber \\
 &  & \mbox{ on the interval} \mbox{} (2^{-n}l,2^{-n}(l+1)) \, \nonumber \\ 
 &   &\mbox{ for} \mbox{} 0\leq l < 2^n, \mbox{} f \mbox{ vanishes elsewhere} \} \, .
\label{eq:wavelets.1}
\end{eqnarray}

\begin{figure}[ht!]
\includegraphics[angle=0,width=0.8\textwidth]{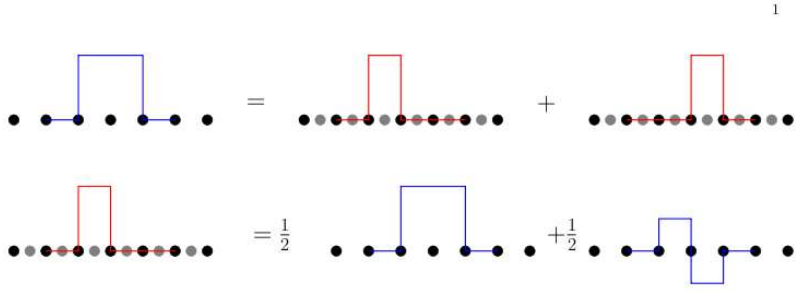}
\caption{Wavelets (bottom) and scaling function (top).
          \label{fig:wavelet-sf}
}
\end{figure}

It is quite obvious that one polynomial of degree $j$ on the interval
$[0,1]$ can be exactly reproduced by two polynomials of degree $j$,
one on the interval $[0,\frac{1}{2}]$ and the other on the interval
$[\frac{1}{2},1]$.  The spaces $V_j^n$ hence fulfills the MRA
condition Eq.~(\ref{eq:mra.1}), and if the polynomial basis is chosen to be
orthogonal, the $V_j^n$ constitutes an orthogonal basis.
\subsection{An example: Simple Haar wavelets}
\begin{figure} 
\begin{center}
\includegraphics[width=0.7\textwidth]{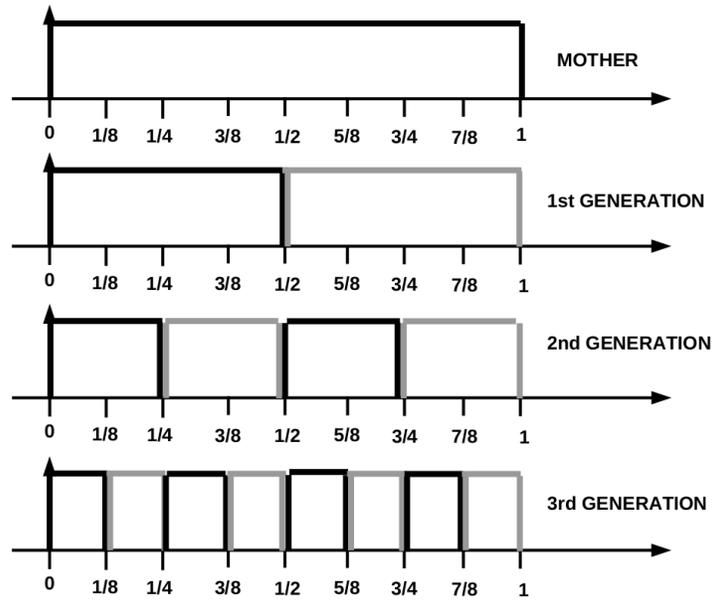}
\end{center}
\caption{Haar scaling functions. \label{fig:Haar_scaling_functions} }
\end{figure}
The basic wavelet ideas that we need can be easily explained using Haar wavelets \cite{H10}.
These are simply the box functions shown in Fig.~\ref{fig:Haar_scaling_functions}.  We begin
with a compact ``mother scaling function,'' in this case the Haar function,
\begin{equation}
  \varphi(x) = \left\{ \begin{array}{ccc} 0 & ; & x>1 \\
                                         1 & ; & 0<x<1 \\
                                         0 & ; & x<0 \end{array} \right. \, .
  \label{eq:mw.1}
\end{equation}
Translations, $\left\{ \varphi_i(x) = \varphi(x-i) \right\}$, of this mother function produces
a crude basis set.  Its relation to the grid of integers is obvious.  Successively more refined
basis sets may be generated by repeated application of the scaling operation consisting of
contracting the functions to half their size in the $x$ direction.  The $k$th generation of
scaling function is given by $\left\{ \varphi_i^{(k)}(x) = \varphi(2^k x-i) \right\}$.
Each generation has a fixed resolution related to an underlying grid with the same resolution.
\begin{figure} 
\begin{center}
\includegraphics[width=0.7\textwidth]{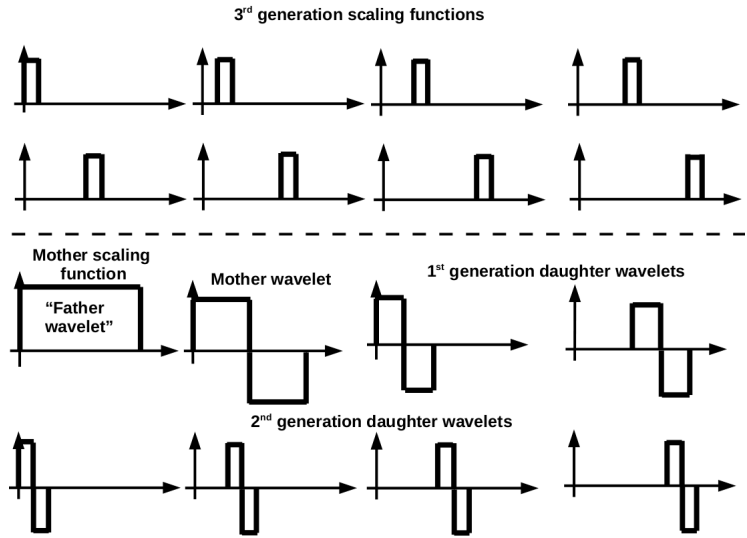}
\end{center}
\caption{Haar scaling functions and the corresponding wavelets. \label{fig:Haar_wavelets} }
\end{figure}
Let us now try to construct a multiresolution basis set.  This is accomplished
by (say) beginning with the third generation wavelets and taking sums and differences
of adjacent functions until the eight third generation scaling functions have been replaced
with the eight wavelets shown in Fig.~\ref{fig:Haar_wavelets}.  Notice how each generation
of daughter wavelets is related to the mother wavelet by scaling,
$\phi_i^{(k)}(x) = \phi(2^k x - i)$.  Notice also how the mother and two generations of daughter wavelets
plus the mother scaling function (occasionally refered to as the ``father wavelet'') constitute
a multiresolution basis set equivalent to the original third generation scaling basis set.
Thus an arbitrary function, $f(x)$, expressible in the original scaling basis,
\begin{equation}
  f(x) = \sum_{i=1}^8 \varphi_i^{(3)}(x) s_i^{(3)} \, ,
  \label{eq:mw.2}
\end{equation}
has the wavelet transform,
\begin{equation}
  f(x) = \varphi_0^{(0)}(x) s_0^{(0)} + \phi_0^{(0)}(x) d_0^{(0)} 
       + \sum_{i=0,1} \phi_i^{(1)}(x) d_i^{(1)} + \sum_{i=0,3} \phi_i^{(2)}(x) d_i^{(2)} \, .
  \label{eq:mw.3}
\end{equation}
Since the basis set is multiresolution, we may choose to add more grid points in some region of
space and go locally to higher order wavelet expansions.  It is also not always necessary to carry
out a full wavelet transform, but rather it may be useful to just carry out a partial transform
giving a linear combination of wavelets with several scaling functions at a time.  The extension
to three dimensions is accomplished by using products of one-dimensional scaling functions and
wavelets.
Haar wavelets are just one type of wavelet basis set.  It happens to be pedagogically useful
but is not particularly useful for computations.
\subsection{Wavelet Basis}
The {\em wavelet space} $W_j^n$ is defined, according to Eq.~(\ref{eq:mra.4}),
as the orthogonal complement of $V_j^n$ in $V_j^{n+1}$.  The
wavelet basis functions of $W_j^n$ are hence piece-wise polynomials
of degree $\leq j$ on each of the two intervals on scale $n+1$ that
overlaps with one interval on scale $n$.  These piece-wise polynomials
are then made orthogonal to a basis of $V_j^n$ and to each other.  The
construction of the wavelet basis follows exactly \cite{A93} where a simple
Gram-Schmidt orthogonalization were employed to construct a basis that
met the necessary orthogonality conditions.

One important property of the wavelet basis is the number of 
vanishing moments.  The $j$-th continuous moment of a function 
$\phi$ is defined as the integrals
\begin{equation}
\mu_j \begin{array}{c} \mbox{def} \\ = \end{array} \int_0^1 x^j \phi(x) dx \, , 
\label{eq:wavelet-basis.1}
\end{equation}
and the function $\phi$ has $M$ vanishing moments if 
$$ \mu_j=0, \quad k = 0, ..., M-1 $$

The vanishing momenets of the {\em wavelet} functions gives information
on the approximation order of the {\em scaling} functions.  If the
wavelet function $\phi$ has $M$ vanishing moments, any polynomial of
order $\leq M-1$ can be exactly reproduced by the scaling function
$\varphi$, and the error in representing an arbitrary function in
the scaling basis is of $M$-th order.  By construction, $x^i$ is in
the space $V_j^0$ for $0 \leq i \leq j$, and since $W_j^0 \perp V_j^0$,
the first $k+1$ moments of $\phi_j^0$ must vanish.
\subsection{The scaling basis}
The construction of the scaling functions is quite straightforward,
$j+1$ suitable 	polynomials are chosen to span any polynomial of degree 
$j$ on the unit interval. The total basis for $V_j^n$  is then obtained by 
appropriate dilation and translation of these functions. 
Of course, any polynomial basis can be used, the simplest of them 
the standard basis $\{1, x, ...,x^j\}$. However, this basis is
not orthogonal on the unit interval. In the following, two choices of 
orthogonal scaling functions will be presented, 
and even though they span exactly the same spaces $V_j^n$ there are some 
important numerical differences between the two. 

In order to construct a set of orthogonal polynomials we could proceed in the
same manner as for the wavelet functions and do a Gram-Schmidt
orthogonalization of the standard basis $\{1,x,...x^j\}$.  If this is
done on the interval $x \in [-1,1]$ we end up with the Legendre
polynomials $\{L_k\}_{k=0}^j$.  These functions are usually normalized such
that $L_k(1) = 1$ for all $j$.  To make the {\em Legendre scaling functions}
$\varphi_k^L$ we transform the Legendre polynomials to the
interval $x \in [0,1]$, and $L^2$ normalize
\begin{equation}
\varphi_k^L (x) = \sqrt{2k+1} L_k (2x-1), \quad x \in[0,1] \, .
\label{eq:scaling-basis.1}
\end{equation}
The basis for the space $V_j^n$ is then made by proper dilation and
translation of $\varphi_k^L$.  
Alpert et al. \cite{A93} presented an alternative set of scaling
functions with interpolating properties.  These
{\em interpolating scaling functions} $\varphi_k^I$ are based on the
Legendre scaling functions $\{\varphi_k^L\}_{k=0}^j$, and the roots
$\{y_k\}_{k=0}^j$ and weights $\{w_k\}_{k=0}^j$ of the Gauss-Legendre
quadrature of order $j+1$, and are organized in the linear combinations
\begin{equation}
\varphi_k^I(x) = \sqrt{w_k}\sum_{i=0}^{j_p} \varphi_i^L (y_k)\varphi_i^L(x), \quad x \in [0,1] \, ,
\label{eq:scaling-basis.2}
\end{equation}
Again the basis of $V_j^n$ is made by dilation and translation of $\phi_k^I$.
The construction of $\varphi_k^I$ gives them the interpolating property
\begin{equation}
\varphi_k^I(y_i) = \frac{\delta_{ki}}{\sqrt{w_i}}\, .
\label{eq:scaling-basis.3}
\end{equation}
which will prove important for numerical efficiency.

A detailed discussion on the properties of interpolating wavelets can be
found in Donoho \cite{D92}, but the case of interpolating wavelets is
somewhat different.  An important property of interpolating wavelets 
is the smoothness of any function represented in this basis.  This
property stems from general Lagrange interpolation.  In the wavelet
case the interpolating property applies within one scaling function vector
only, which means that functions represented in this basis can be
discontinous in any merging point between the different translations on 
any scale.  
\subsection{Interpolating scaling functions}
Since the general introduction to wavelets were already made,
we will now concencentrate our
description on the level 3 interpolating scaling function (ISF)  
introduced by Deslauriers and Dubuc, and described in detail
in Ref. \cite{LAE98}.
Its main advantage is that it is fast
and easy to perform nonlinear operations on functions
represented in this basis, as long as the operation is local
in shape.  It also represents 3rd order polynomials exactly
which means that it behaves very smoothly.

We introduced the projection operator $P^n$ that projects
an arbitrary function $f(x)$ onto the basis $\{\varphi_{j,l}^n\}$
of the scaling space $V^n$ (in the remaining of this text
the subscript $k$ of the scaling and wavelet spaces
will be omitted, and it will always be assumed that we
are dealing with a $k$th order polynomial basis.)
\begin{equation}
  f(x)\approx P^n f(x) \begin{array}{c} \mbox{def} \\ = \end{array} f^n(x) = \sum_{l=0}^{2^n-1} \sum_{j=0}^k s_{j,l}^{n,f} \varphi_{j,l}^{n} (x) \, ,
  \label{eq:isf.1}
\end{equation}
where the expansion coefficients $s_{j,l}^{n,f}$, the so-called
{\em scaling } coefficients, are obtained by the usual integral
\begin{equation}
  s_{j,l}^{n,f} \begin{array}{c} \mbox{def} \\ = \end{array} 
  \langle f, \varphi_{j,l}^n \rangle = \int_0^1 f(x) \varphi_{j,l}^{n} (x) dx \, ,
  \label{eq:isf.2}
\end{equation}
If this approximation turns out to be too crude, we double our basis set
by increasing the scale and perform the projection $P^{n+1}$.  This
can be continued until we reach a scale $N$ where we are satisfied with the
overall accuracy of $f^N$ relative to the true function $f$.

In a perfect world, the projection in Eq.~(\ref{eq:isf.2}) could be
done exactly, and the accuracy of the projection would be
independent of the choice of polynomial basis.  In the real world the
projections are done with Gauss-Legendre quadrature and the expansion
coefficients $s_{j,l}^{n,f}$ of $f(x)$ are obtained as
\begin{eqnarray}
  s_{j,l}^{n,f} & = & \int_{2^{-nl}}^{2^{-n(l+1)}} f(x) \varphi_{j,l}^n (x) dx \, \nonumber \\
  & = & 2^{-n/2} \int_0^1 f(2^{-n} (x+l)) \varphi_{j,0}^0 (x) dx \, \nonumber \\
  & \approx & 2^{-n/2} \sum_{q=0}^{k_q-1} w_q f(2^{-n}(y_q+l)) \varphi_{j,0}^0 (y_q) \, 
  \label{eq:isf.3}
\end{eqnarray}
where $\{w_q\}_{q=0}^{k_q-1}$ are the weights and $\{y_q\}_{q=0}^{k_q-1}$
the roots of the Legendre polynomial $L_{k_q}$ used in $k_q$th order
quadrature.

By approximating this integral by quadrature we will of course not
obtain the exact expansion coefficients.  However, it would be nice if we
could obtain the exact coefficients whenever our basis is flexible
enough to reproduce the function exactly, that is if $f(x)$ is a
polynomial of degree $\leq k$. The Legendre quadrature holds a $(2k-1)$
rule which states that the $k$-order quadrature is exact whenever the
integrand is a polynomial of order $2k-1$.  By choosing $k_q=k+1$ order
quadrature we will obtain the exact coefficient whenever $f(x)$ is a
polynomail of degree $\leq (k+1)$ when projecting on the basis
of $k$-order Legendre polynomials.

In the multidimensional case the expansion coefficients are
given by multidimensional quadrature
\begin{equation}
  s_{jl}^{nf} = 2^{-dn/2} \sum_{q_1=0}^k \sum_{q_2=0}^k ... \sum_{q_d=0}^k f(2^{-n}(y_q +l)) \Pi_{i=1}^d w_{q_i} \varphi_{j_p,0}^0 (y_{q_i}) \, ,
  \label{eq:isf.4}
\end{equation}
using the following notation for the vector of quadrature roots
\begin{equation}
  y_q \begin{array}{c} \mbox{def} \\ = \end{array}  (y_{q_1}, y_{q_2},...,y_{q_d}) \, ,
  \label{eq:isf.5}
\end{equation}
This quadrature is not very efficient in multiple dimensions since
the number of terms scales as $(k+1)^d$.  However, if the function $f$
is separable and can be written $f(x_1,x_2,...,x_d) = f_1(x_1) f_2(x_2) ...f_d(x_d)$, Eq.~(\ref{eq:isf.4}) can be simplified to
\begin{equation}
  s_{jl}^{nf} = 2^{-dn/2} \Pi_{i=1}^d \sum_{q_i=0}^k f_i (2^{-n}(y_{q_i} +l_i))w_{q_i} \varphi_{j_i,0}^0(y_{q_i}) \, ,
  \label{eq:isf.6}
\end{equation}
which is a product of small summations and scales only as $d(k+1)$.

The Legendre polynomials show very good convergence for polynomial
functions $f(x)$, and are likely to give more accurate projections.
However, most interesting functions $f(x)$ are not simple polynomials,
and the accuracy of the Legendre scaling functions versus a general
polynomial basis might not be very different.

By choosing the quadrature order to be $k+1$ a very important property
of the Interpolating scaling functions emerges, stemming from the
specific construction of these functions Eq.~(\ref{eq:scaling-basis.2}),
and the use of the $k+1$ order quadrature roots and weights.  The
interpolating property Eq.~(\ref{eq:scaling-basis.3}) inserts a
Kronecker delta whenever the scaling function is evaluated in a
quadrature root, which is exactly the case in the quadrature sum.
This reduces Eq.~(\ref{eq:isf.3}) to
\begin{equation}
  s_{jl}^{n,f} = \frac{2^{-n/2}}{\sqrt{w_j}} f(2^{-n}(x_j+l)) \, ,
  \label{eq:isf.7}
\end{equation}
which obviously makes the projection $k+1$ times more efficient.

In multiple dimensions this property becomes even more important,
since it effectively removes all the nested summations in Eq.~(\ref{eq:isf.4}) and leaves only one term in the projection
\begin{equation}
  s_{jl}^{nf} = f(2^{-n}(y_j +l)) \Pi_{i=1}^d \frac{2^{-n/2}}{\sqrt{w_{j_i}}} \, ,
  \label{eq:isf.8}
\end{equation}
This means that in the Interpolating basis the projection is equally effective
regardless of the separability of the function $f$.

\section{Density Functional Theory}
\label{sec:dft}

A method to resolve the electronic structure is by using variational principle
\begin{equation}
  \label{eq:variation-principle}
  E[\Psi] = \frac{\langle\Psi|\hat H|\Psi\rangle}{\langle\Psi|\Psi\rangle} \, ,
\end{equation}
Where $\langle\Psi|\hat H|\Psi\rangle = \int \, d{\bf r} \Psi^*({\bf r}) \hat H \Psi({\bf r})$, $\Psi$ denotes
the electronic wavefunction and $\hat H$ the Hamiltonian.  The energy computed
from a guess $\Psi$ is an upper bound to the true ground
state energy $E_0$.  Full minimization of the functional $E[\Psi]$ will
give the true ground state $\Psi^{gs}$ and energy $E_0 = E[\Psi^{gs}]$.

Density-functional theory states that the many electron problem
can be replaced by an equivalent set of self-consistent one-electron
equations, the Kohn-Sham equations
\begin{equation}
  \label{eq:ks}
  \hat h\psi_i^{\sigma}({\bf r}) = \left ( -\frac{1}{2} \nabla^2 + \hat v_{pp}({\bf r}) 
  + \hat v_{H}({\bf r}) +\hat v_{xc}^{\sigma} ({\bf r})\right ) \psi_i^{\sigma} ({\bf r}) 
  = \epsilon_i^{\sigma} \psi_i^{\sigma} ({\bf r}) \, .
\end{equation}
The eigenfunctions $\psi_i^{\sigma}$ are the one-electron wavefunctions that correspond to the
minimum of the Kohn-Sham energy functional.  In these
wavefunctions, {\it i} is the orbital index and $\sigma$ denotes the spin, which
can be either up $\uparrow$ or down $\downarrow$ (spin $\alpha$ or $\beta$.)

The Hamiltonian $\hat H$ consists of four different parts: a part related to the 
kinetic energy of the electrons, the pseudopotentials $\hat v_{psp}$, the
Hartree potential $\hat v_{H}$ and the exchange correlation potential $\hat v_{xc}$.
The interaction of the positively charged nuclei with the electrons is 
described using the pseudopotential $\hat v_{psp}$ instead of using the
full Coulombic potential.  The pseudopotential usually consists of both
a local and a non-local part
\begin{equation}
  \label{eq:psp}
  \hat v_{psp} ({\bf r}) = v_{loc} (r) + \sum_{l} |l\rangle \hat v_l (r, r')\langle l| \, .
\end{equation}
The Hartree potential $\hat v_{H}$ describes the interaction between
electrons and is given by
\begin{equation}
  \label{eq:hartree}
  \hat v_{H} ({\bf r}) = \int d{\bf r'} \frac{\rho_{\uparrow}({\bf r})+\rho_{\downarrow}({\bf r'})} 
  {|{\bf r} - {\bf r'}|} \, .
\end{equation}
Finally, the exchange correlation potential $\hat v_{xc}$ describes the nonclassical interaction
between the electrons and is given by the functional derivative of an exchange
correlation energy functional
\begin{equation}
  \label{eq:xc}
  \hat v_{xc}^{\sigma} ({\bf r}) = \frac{\delta E_{xc} (\rho_{\uparrow}, \rho_{\downarrow})}
  {\delta\rho_{\sigma}({\bf r})} \, .
\end{equation}
In these equations $\rho^{\sigma}$ is the electron spin density, defined as
\begin{equation}
  \label{eq:density}
  \rho_{\sigma}({\bf r}) = \sum_{i} n_i^{\sigma} |\psi_i^{\sigma} ({\bf r}) |^2 \, ,
\end{equation}
where $n_i^{\sigma}$ is the occupation number, i.e. the number of electrons in orbital {\it i}.
In case of LDA (which we use throughtout this chapter) where there is no longer a distinction between 
spin up and spin down, orbitals
can contain at most two electrons.

\section{Time-Dependent Density Functional Theory}
\label{sec:tddft}
This section contains a brief review of the
basic formalism of TD-DFT which is already well-known from the literature \cite{RG84}.
The time-dependent single particle Kohn-Sham equations are,
\begin{equation}
  \left ( -\frac{1}{2} \nabla^2 + v_{eff}[\rho]({\bf r}, t)\right) \psi_{i \sigma}({\bf r},t) = i \frac{\partial}{\partial t}\psi_{i \sigma}({\bf r},t)
  \label{eq:TD-KS}
\end{equation}
Here, the wave functions $\psi_i({\bf r},t)$ and $v_{eff}[\rho]({\bf r}, t)$ explicitly depend on time,
whereas,
\begin{equation}
  v_{eff}[\rho]({\bf r},t) = \sum_a v_{ion}({\bf r} - {\bf R}_a) + \int \frac{\rho[{\bf r'}, t]}{|{\bf r} -{\bf r'}|} d {\bf r'} 
  + v_{xc}[\rho]({\bf r},t) \, .
  \label{eq:v-eff}
\end{equation}
Using  the adiabatic approximation (AA), (which is local in time) 
\begin{eqnarray}
  v_{xc}[\rho]({\bf r},t) & \approx & \frac{\delta E_{xc}[\rho]}{\delta \rho({\bf r})} \nonumber \\
  \frac{\delta v_{xc}[\rho]({\bf r},t)}{\delta \rho({\bf r'},t)} & \approx & 
  \delta (t-t') \frac{\delta^2 E_{xc}[\rho]}{\delta \rho({\bf r})\delta \rho({\bf r'})} \, ,
  \label{eq:v-xc}
\end{eqnarray}
and using the LDA,
\begin{equation}
  E_{xc}[\rho] = \int \rho({\bf r}) \epsilon_{xc} (\rho({\bf r})) d {\bf r} \, .
  \label{eq:exc}
\end{equation}

The method that we  wish to use here is Casida's approach \cite{C95}.
This section explains the deriving equations of  linear-response (LR) TD-LDA  method. 

The time-dependent perturbation to the external potential can be written as,
\begin{equation}
  \delta v_{eff}[\rho]({\bf r},t) = \delta v_{appl}({\bf r},t) + \delta v_{SCF}[\rho]({\bf r},t)
  \label{eq:td v-eff}
\end{equation}
where,
\begin{equation}
  v_{SCF}[\rho]({\bf r},t) = \int \frac{\rho({\bf r'},t)}{|{\bf r} - {\bf r'}|} d {\bf r'} + v_{xc}[\rho]({\bf r'},t)
  \label{eq:v-scf}
\end{equation}
The LR of the density matrix (DM) can be written in terms of generalised susceptibility $\chi$ as below, 
\begin{equation}
  \delta {\bf P}_{ij\sigma}(\omega) = \sum_{kl\tau} \chi_{ij\sigma,kl\tau} (\omega) \delta v_{kl\tau}^{eff}(\omega)
  \label{eq:lr rho}
\end{equation}
After some mathematical steps, one can end-up with the sum-over-states (SOS) representation of $\chi$, 
\begin{equation}
  \chi_{ij\sigma,kl\tau}(\omega) = \delta_{ik} \delta_{jl} \delta_{\sigma\tau} \frac{\lambda_{lk\tau}}{\omega-(\omega_{lk\tau})}
  \label{eq:sos chi}
\end{equation}
where $\lambda_{lk\tau}= n_{l\tau}-n_{k\tau}$ the difference in occupation numbers and $\omega_{lk\tau} = \epsilon_{k\tau} -\epsilon_{l\tau}$
 the difference between the eigenvalues of $l$th and $k$th states.
In the basis of Kohn-Sham orbitals\{$\psi_{i\sigma}$\}, we can re-write the LR-DM equation as,
\begin{equation}
  \delta {\bf P}_{ij\sigma} (\omega) = \frac{\lambda_{ji\sigma}}{\omega - \omega_{ji\sigma}} 
  \left [\delta v_{ij\sigma}^{appl}(\omega) + \delta v_{ij\sigma}^{SCF}(\omega) \right ]
  \label{eq:lr rho ks}
\end{equation}
Now the term $\delta v^{SCF}$ is complicated because it itself depends on the response of the DM.
\begin{equation}
  \delta v_{ij\sigma}^{SCF} (\omega) = \sum {\bf K}_{ij\sigma,kl\tau} \delta {\bf P}_{kl\tau}(\omega)
  \label{eq:delta v-scf}
\end{equation}
Where,
\begin{equation}
  {\bf K}_{ij\sigma,kl\tau} = \frac{\partial v_{ij\sigma}^{SCF}}{\partial {\bf P}_{kl\tau}}
  \label{eq:coupling matrix}
\end{equation}
whose integral form is,
\begin{equation}
  {\bf K}_{ij\sigma,kl\tau} = 
  \int \int \psi_{i\sigma}^{*}({\bf r}) \psi_{j\sigma}({\bf r}) 
  \left [ \frac{1}{|{\bf r} - {\bf r'}|} + \frac {\delta^2 E_{xc}[\rho]}{\delta \rho_\sigma({\bf r})\delta \rho_\tau ({\bf r'})} \right ] 
  \psi_{k\tau}({\bf r'}) \psi_{l\tau}^{*}({\bf r'}) d {\bf r} d {\bf r'}
  \label{eq:integral k}
\end{equation}
If the response is due to a real spin independent external perturbation, $\delta v^{appl}$, then we can restrict ourselves
to the real density response and the coupling matrix is symmetric.  

After some algebra, the real parts of the DM elements $ \Re \delta {\bf P}(\omega)$ can be given as,
\begin{equation}
   \sum_{kl\tau} \left [ \frac{\delta_{ik},\delta_{jl}\delta_{\sigma\tau} }{\lambda_{kl\tau}\omega_{kl\tau}} (\omega^2 - \omega_{kl\tau}^2) -  2 {\bf K}_{ij\sigma,kl\tau}\right ] 
   \Re (\delta {\bf P}_{kl\tau})(\omega) = \delta v_{ij\sigma}^{appl}(\omega)
   \label{eq:170}
\end{equation}
Here the real part of  $ \Re \delta {\bf P_\sigma}(\omega)$ means the Fourier transform of the real part of  $ \Re \delta {\bf P_\sigma}(t)$.
Thus the real part of the first-order DM obtained from the solution of the above linear equations gives access to the
frequency-dependent polarizabilities. This leads to the following eigenvalue equation from which the excitation energies
and oscillator strengths can be obtained.
\begin{equation}
  {\bf \Omega} {\vec F}_I = {\it\omega}_I^2 {\vec F}_I \, ,
  \label{eq:casida}
\end{equation}
where,
\begin{equation}
  {\bf \Omega}_{ij\sigma,kl\tau} = \delta_{ik}\delta_{jl}\delta_{\sigma\tau}\omega_{kl\tau}^2 + 2 \sqrt{\lambda_{ij\sigma}\omega_{ij\sigma}} {\bf K}_{ij\sigma,kl\tau}
  \sqrt{\lambda_{kl\tau}\omega_{kl\tau}}
  \label{eq:method.180}
\end{equation}
Here, the desired excitation energies are equal to ${\bf \omega_I}$ and the oscillator strengths $f_I$ are obtained from the
 eigenvectors ${\vec F}_I$. The frequency-dependent polarizability is directly realted to vertical excitation energies,
oscillator strength and transition dipole moments $\mu_I$,
\begin{equation}
  \alpha(\omega) = \sum_I \frac{f_I}{{\it\omega_I}^2-\omega^2} = \frac{2}{3} \sum_I \frac{{\bf \omega_I}\mu_I^2 }{{\bf \omega_I}^2 - \omega^2}
  \label{eq:oscillator strength}
\end{equation}


\section{Krylov Space Methods}
\label{sec:krylov}

The methods described in this article involve solving very large eigenvalue problems.
One of these is the matrix form of the Kohn-Sham orbital equation Eq.~(\ref{eq:ks})
while the other is the LR-TD-DFT equation Eq.~(\ref{eq:casida}).  The first is very
large because the wavelet basis set is very large while the other is very large because
it is of the order of the number of unoccupied orbitals times the number of unoccupied
orbitals on each side.  It is important to realize that special methods must be and are
being used to solve these very large eigenvalue problems.  In particular, {\sc BigDFT}
make use of the block Davidson variant of the Krylov space method to solve the Kohn-Sham
equation while {\sc BigDFT} and most other codes solving the LR-TD-DFT equation 
Eq.~(\ref{eq:casida}) also make use of the the block Davidson method.  Krylov methods
and the block Davidson method are briefly described in this section.

Krylov space methods may be traced back to a paper in the early 1930s written by 
the Russian mathematician Alexei Nikolaevich Krylov.  The main idea is that to solve
a matrix problem involving a matrix ${\bf A}$, it is frequently never actually necessary
to construct the matrix ${\bf A}$ because iterative solutions only require a reasonable
first guess followed by repeated action of the {\em operator} $\hat{A}$ on a vector.
A number of such methods are known with Lanczos diagonalization and the discrete inversion
in the iterative subspace (DIIS) \cite{P80} as particlarly well-known examples in theoretical  
chemical physics.  Given a vector $\vec{x}$, the Krylov space of order $r$ is given by,
\begin{equation}
  {\cal K}_r({\bf A},\vec{x}) = \mbox{span} \left\{ \vec{x}, {\bf A} \vec{x}, {\bf A}^2 \vec{x}, \cdots ,
  {\bf A}^r \vec{x} \right\} \, .
  \label{eq:krylov.1}
\end{equation}

The Davidson diagonalization method \cite{D75} for solving the matrix eigenvalue problem
\begin{equation}
  {\bf A} \vec{x} = a \vec{x} \, , 
  \label{eq:krylov.2}
\end{equation}
is deceptively simple. Suppose that we want the lowest eigenvalue and eigenvector and we have
an intial guess vector, $\vec{x}^{(0)}$.  Then we can always write,
\begin{equation}
  \vec{x} = \vec{x}^{(0)} + \delta \vec{x} \, ,
  \label{eq:krylov.3}
\end{equation}
is the component of the exact solution which is orthogonal to the intial gues vector.
Simple algebra then gives a formula highly reminiscent of Rayleigh-Schr\"odinger 
perturbation theory but exact,
\begin{equation}
  \delta \vec{x}  = \left[ {\bf Q} (a {\bf 1} - {\bf A}) {\bf Q} \right]^{-1} 
  \left( {\bf A} - a {\bf 1} \right) \vec{x}^{(0)} \, ,
  \label{eq:krylov.4}
\end{equation}
where,
\begin{equation}
  {\bf Q} = {\bf 1} - \vec{x}^{(0)} \vec{x}^{(0)\dagger} \, ,
  \label{eq:krylov.5}
\end{equation}
projects onto the subspace orthogonal to the guess vector.  Solving Eq.~(\ref{eq:krylov.4})
requires us to overcome two difficulties.  The first is that we need a guess for the 
eigenvalue $a$, but this is easily remedied by taking 
$a^{(0)}=\vec{x}^{(0)\dagger} {\bf A} \vec{x}^{(0)}/\vec{x}^{(0)\dagger}\vec{x}^{(0)}$ and
then iterating.  The problem greater problem is to invert the matrix 
$\left[ {\bf Q} (a {\bf 1} - {\bf A}) {\bf Q} \right]$.  It actually turns out that 
a highly accurate inversion is not really needed (and actually can cause some problems.)
Instead it is better just to replace this difficult inversion with,
\begin{equation}
  \delta \vec{x}  = \left( a {\bf 1} - {\bf D} \right)^{-1} 
  \left( {\bf A} - a {\bf 1} \right) \vec{x}^{(0)} \, ,
  \label{eq:krylov.6}
\end{equation}
where ${\bf D}$ is some diagonal matrix, hence easy to invert.  Orthogonalizing $\delta \vec{x}$
to $\vec{x}^{(0)}$ and normalizing produces $\vec{x}^{(1)}$, which is the next basis vector in
our Krylov space.  Setting up and diagonalizing the 2 $\times$ 2 matrix of ${\bf A}$ in this
basis and taking the lowest eigenvalue gives us the next estimate $a^{(1)}$.  If application of
Eq.~(\ref{eq:krylov.4}) is close to zero then we have solved the eigenvalue problem 
Eq.~(\ref{eq:krylov.2}), otherwise we have a new $\delta \vec{x}$ with which to generate
$\vec{x}^{(2)}$ and so on and so forth until convergence.  The block Davidson method
\cite{MRD91} extends the Davidson method to the lowest several eigenvalues and eigenvectors.

Davidson diagonalization works well when started from a reasonably good initial guess, otherwise
the Lanczos method may be advantageous.  One of the most recent incarnations of the Lanczos
method is the Liouville-Lanczos method for solving the LR-TD-DFT problem 
\cite{WSGB06,RGSB08,SGM+10,MGRB11}.


\section{Numerical Implementation of DFT in {\sc BigDFT}}
\label{sec:bigdft-dft}
Computational physics/chemistry is the transformation and
implementation of scientific theory into efficient algorithms
which requires both theoretical and experimental skill.  The
transformation of a new theory into an efficient algorithm
requires understanding of programming concepts, mathematical and
physical intuition and theoretical insight, whereas the production
of the computer code is much like experimentation, requiring debugging,
testing and organisation to yield a highly efficient product.
It is also an adaptation of new scientific theory into
computer code exploiting the advances in compiler, programming language
and hardware technology.  The aim is to afford an algorithm to enable
efficient computation, portability of code, ease of adaptability and 
to document the science.  To afford such an algorithm requires an intuitive
understanding of the physics to be implemented, much experimentation with
optimisation and debugging of the developing code, a suitable choice of 
programming language, as well as a basic overview of the nature of the
platforms for which the code is intended.

The Kohn-Sham scheme of DFT greatly reduces the complexity of ground state
electronic structure calculations by recasting the many-body
problem into a (self-consistent) single-particle problem. For real
atomistic systems, however, the KS equations are still difficult to solve and 
further approximate techniques are required.
In general it is important, though, that these approximations can be
controlled in such a way that the associated error does not exceed the error
introduced by the xc-functional.
While DFT accounts for approximately 90$\%$ of all quantum chemical
calculations being performed, the sometimes unpredictable nature of results
and the inability to systematically improve the quality of calculation may mean
that a place for the conventional correlated techniques remains in the
quantum chemist's tool kit.  
In this work the detailed description of DFT program {\sc BigDFT}
has been given. {\sc BigDFT} \cite{BigDFT} has been developed 
as an European project (FP6-NEST) from 2005 to 2008,
and is a wavelet-based pseudopotential implementation of DFT and TD-DFT.
For complimentary purpose, Gaussian based quantum chemistry DFT code {\sc deMon2k} \cite{deMon2k}
is also used but we are not going to discuss the numerical implementation of {\sc deMon2k} here and we restrict ourselves to use {\sc deMon2k} for revalidating our recent implementation of LR-TD-DFT in {\sc BigDFT}.  
However in the following sections, we are going to recast how the
fundamental computational operations were performed in {\sc BigDFT}.
\subsection{Daubechies Wavelets}
Before embarking on our own endeavours, we should make some reference to related
work.  First, it should be acknowledged that a considerable amount of work
has been done already in pursuit of a wavelets in the electronic
structure calculations \cite{GDNGB06,GDG07,GAGD+08,GODM+09,DG11,GVODGM11}. The object of using  wavelets as basis set is to
associate an expansion coefficient to each of the piece-wise
wavelets.  The expansion coefficients are free to vary from one wavelet
function to the next.  This feature enables wavelets as highly localised
continuous functions of a fractal nature that have finite supports.
The Daubechies wavelets have no available analytic forms, and
they are not readily available in sampled versions. They are defined effectively
by the associated dilation coefficients.  These express a wavelet in high
resolution and a scaling function in the low resolution--which
has the same width and which stretches to zero--as a linear combination
of the more densely packed and less dispersed scaling functions
that form a basis for the two resolution level in combination.

The fact that the Daubechies wavelets are known only via their dilation
coefficients is no implediment to the discrete wavelet transform.  This
transform generates the expansion coefficients associated with the wavelet
decomposition of a data sequence.  In this perspective, the dilation
coefficients of the wavelets and of the associated scaling functions are
nothing but the coefficients of a pair of quadrature mirror filters that
are applied succesively.

\begin{figure}[h!]
  \begin{center}
  \includegraphics[angle=0,width=0.5\textwidth]{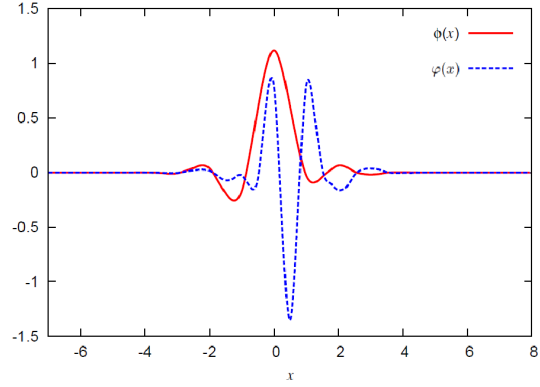}
  \end{center}
  \caption{Daubechies scaling function $\phi(x)$ and wavelet $\varphi(x)$ of order  16. \label{fig:daubechies} }
\end{figure}

As described above Daubechies family consists of two fundamental functions: the
scaling function $\phi(x)$ and the wavelet $\varphi(x)$ (see
Fig. \ref{fig:daubechies}.) The full basis set can be
obtained from all translations by a certain grid spacing {\it h}
of the scaling and wavelet functions centered at the origin.
These functions
satisfy the fundamental defining (refinement) equations,
\begin{eqnarray}
  \phi(x) & = & \sqrt{2} \sum_{j=1-m}^m h_j \phi(2x-j) \nonumber \, , \\
  \varphi(x) & = & \sqrt{2} \sum_{j=1-m}^m g_j \phi(2x-j) \, .
  \label{eq:daub.1}
\end{eqnarray}
which relate the basis functions on a grid with spacing $ h$ and
another one with spacing $ h/2$.
The coefficients, h$_j$ and g$_j$, consitute the so-called ``filters"
which define the wavelet family of order $m$.  These coefficients
satisfy the relations, $\sum_j h_j = 1$ and $g_j = (-1)^j h_{-j+1}$.
Eq.~(\ref{eq:daub.1}) is very important since it means that
a scaling-function basis defined over a fine grid of spacing $h/2$
may be replaced by combining a scaling-function basis over a
coarse grid of spacing $h$ with a wavelet basis defined over
the fine grid of spacing $h/2$.  This then gives us the liberty
to begin with a coarse description in terms of scaling functions
and then add wavelets only where a more refined description is needed.
In principle the refined wavelet description may be further refined
by adding higher-order wavelets where needed.  However in {\sc BigDFT}
we restricted ourselves to just two levels: coarse and fine associated
respectively with scaling functions and wavelets.

For a three-dimensional description, the simplest basis set is
obtained by a tensor product of one-dimensional basis functions.
For a two resolution level description, the
coarse degrees of freedom are expanded by a single three dimensional
function, $\phi_{i_1,i_2,i_3}^0(\bf r)$, while the fine degrees of freedom
can be expressed by adding  another seven basis functions, $\phi_{j_1,j_2,j_3}^{\nu}(\bf r)$,
which include tensor products with one-dimensional wavelet functions.
Thus, the Kohn-Sham wave function $\psi(\bf r)$ is of the form
\begin{equation}
  \psi({\bf r}) = \sum_{i_1,i_2,i_3} c_{i_1,i_2,i_3}^0 \phi_{i_1,i_2,i_3}^0({\bf r}) 
  +\sum_{j_1,j_2,j_3}\sum_{\nu_1}^7 c_{j_1,j_2,j_3}^{\nu} \phi_{j_1,j_2,j_3}^{\nu} ({\bf r}) \, .
\label{eq:daub.2}
\end{equation}
The sum over {\it{i$_1$,i$_2$,i$_3$}} runs over all the grid points contained
in the low-resolution regions and the sum over {\it{j$_1$,j$_2$,j$_3$}} runs over all
the points contained in the (generally smaller) high resolution regions.
 Each wave function is then
described by a set of coefficients $\{c_{j_1,j_2,j_3}^{\nu}\},\nu=0,...,7$.
Only the nonzero
scaling function and wavelet  coefficients are stored. The data is thus
compressed.  The basis set being orthogonal, several operations
such as scalar products among different orbitals and between orbitals
and the projectors of the nonlocal pseudopotential can be directly
carried out in this compressed form.
\subsection{Treatment of kinetic energy}
The matrix elements of the kinetic energy operator among the basis functions
of our mixed representation (i.e., scaling functions with scaling functions,
scaling function with wavelets and wavelets with wavelets) can be calculated
analytically~\cite{B92}. For simplicity, let us illustrate the
application of the kinetic energy operator onto
a wavefunction  $\psi$ that is only expressed in terms of scaling functions.
$$ \psi(x,y,z) = \sum_{i_1,i_2,i_3}\!\!\! s_{i_1,i_2,i_3} 
    \phi(x/h-i_1)\, \phi(y/h-i_2)\, \phi(z/h-i_3) $$
The result of the application of the kinetic energy operator on this wavefunction, projected to the original scaling function space, has the expansion coefficients
\begin{eqnarray}
  \hat{s}_{i_1,i_2,i_3}=-\frac{1}{2h^3}\int  \phi(x/h-i_1)\, \phi(y/h-i_2)\, 
  \phi(z/h-i_3)\times\nonumber\\ 
  \times\nabla^2 \psi(x,y,z) \rm d x \rm d y \rm d z\;.  \nonumber
\end{eqnarray}
Analytically the coefficients $s_{i_1,i_2,i_3}$ and $\hat{s}_{i_1,i_2,i_3}$ are related by a convolution
\begin{equation} 
 \label{eq:kiner}
 \hat{s}_{i_1,i_2,i_3} =\frac{1}{2} \sum_{j_1,j_2,j_3} K_{i_1-j_1,i_2-j_2,i_3-j_3} s_{j_1,j_2,j_3} 
\end{equation}
where
\begin{equation} 
  \label{eq:fprod}
  K_{i_1,i_2,i_3}=T_{i_1} T_{i_2} T_{i_3},
\end{equation}
where the coefficients $T_i$ can be calculated analytically via an eigenvalue equation:
\begin{eqnarray*}
   T_i & = &  \int  \phi(x) \frac{\partial^2}{\partial x^2} \phi(x-i) dx   \\
   & = & \sum_{\nu,\mu} 2 h_{\nu} h_{\mu} \int
      \phi(2x-\nu) \frac{\partial^2}{\partial x^2} \phi(2x-2i-\mu) dx  \\
   & = & \sum_{\nu,\mu} 2 h_{\nu} h_{\mu} 2^{2-1} \int
            \phi(y-\nu) \frac{\partial^2}{\partial y^2} \phi(y-2i-\mu) dy  \\
   & = & \sum_{\nu,\mu} h_{\nu} h_{\mu} 2^2 \int
        \phi(y) \frac{\partial^2}{\partial y^l} \phi(y-2i-\mu+\nu) dy  \\
   & = & \sum_{\nu,\mu} h_{\nu} h_{\mu} \: 2^2 \: T_{2i-\nu+\mu}
\end{eqnarray*}
Using the refinement equation Eq.~(\ref{eq:daub.1}), the values of the $T_i$ can be calculated analytically, from a suitable eigenvector of a matrix derived from the wavelet filters~\cite{B92}. 
For this reason the expression of the kinetic energy operator is \emph{exact} in a given 
Daubechies basis.

Since the 3-dimensional kinetic energy filter $K_{i_1,i_2,i_3}$
is a product of three one-dimensional filters Eq.~(\ref{eq:fprod}) the convolution in Eq.~(\ref{eq:kiner}) can be evaluated with $3 N_1 N_2 N_3 L$ operations for a three-dimensional grid of $ N_1 N_2 N_3 $ grid points.
$L$ is the length of the one-dimensional filter which is 29 for our Daubechies family.
The kinetic energy can thus be evaluated with linear scaling with respect to the
number of nonvanishing expansion coefficients of the wavefunction.
This statement remains true for a mixed scaling function-wavelet basis where we have both nonvanishing $s$ and $d$ coefficients and
for the case where the low and high resolution regions cover only parts
of the cube of $N_1 N_2 N_3$ grid points.
\subsection{Treatment of local potential energy} \label{potentialsection}
In spite of the striking advantages of Daubechies wavelets the initial exploration
of this basis set~\cite{TW97} did not lead to any algorithm that would be useful for
practical electronic structure calculations. This was due to the fact that an accurate evaluation
of the local potential energy is difficult in a Daubechies wavelet basis.

By definition, the local potential $v(\mathbf r)$ can be easily known
on the nodes of the uniform grid of the simulation box.
Approximating a potential energy matrix element $v_{i,j,k;i',j',k'}$
$$ v_{i,j,k;i',j',k'} =
\int \rm d \mathbf{r} \phi_{i',j',k'}({\bf r})  v({\bf r}) \phi_{i,j,k}({\bf r}) $$
by
$$ v_{i,j,k;i',j',k'} \approx \sum_{l,m,n}  \phi_{i',j',k'}({\bf r}_{l,m,n})  v({\bf r}_{l,m,n}) \phi_{i,j,k}({\bf
r}_{l,m,n}) $$
gives an extremely slow convergence rate with respect to the number of grid points
used to approximate the integral because a single scaling function is not very smooth,
i.e., it has a rather low number of continuous derivatives.
A. Neelov and S. Goedecker~\cite{NG06} have shown that one should not
try to approximate a single matrix element as accurately as possible
but that one should try instead to approximate directly the
expectation value of the local potential. The reason for this strategy
is that the wavefunction expressed in the Daubechy basis is smoother than a single Daubechies basis function.
A single Daubechies scaling function of order 16 (i.e., the  corresponding wavelet has  8 vanishing moments) has only 2  continuous derivatives. More precisely its index of  H\"{o}lder continuity is  about 2.7 and the Sobolev space regularity  with respect to $p=2$ is about 2.91~\cite{LS00}.
A single Daubechies scaling function of order 16 has only 4 continuous derivatives.
By suitable linear combinations of Daubechies 16 one can however exactly represent
polynomials up to degree 7, i.e., functions that have 7 non-vanishing continuous derivatives.
The discontinuities get thus canceled by taking suitable linear combinations.
Since we use pseudopotentials, our exact wavefunctions are analytic and can
locally be represented by a Taylor series. We are thus approximating functions
that are approximately polynomials of order 7 and the discontinuities nearly cancel.

Instead of calculating the exact matrix elements we therefore use
matrix elements with respect to a smoothed version $\tilde{\phi}$ of the
Daubechies scaling functions.
\begin{eqnarray}
 v_{i,j,k;i',j',k'} & \approx  & \sum_{l,m,n}  \tilde{\phi}_{i',j',k'}({\bf r}_{l,m,n})  v({\bf r}_{l,m,n}) \tilde{\phi}_{i,j,k}({\bf r}_{l,m,n}) \nonumber \\
  & = & \sum_{l,m,n}  \tilde{\phi}_{0,0,0}({\bf r}_{l-i',m-j',n-k'})  V({\bf r}_{l,m,n}) \tilde{\phi}_{0,0,0}({\bf r}_{l-i,m-j,n-k}) 
  \label{eq:vmag}
\end{eqnarray}
where the smoothed wave function is defined by
$$ \tilde{\phi}_{0,0,0}({\bf r}_{l,m,n})=\omega_l\omega_m\omega_n$$
and $\omega_l$ is the ``magic filter''.
Even though Eq.~(\ref{eq:vmag}) is not a particulary good approximation for a single matrix
element it gives an excellent approximation for the expectation values of the local potential energy
$$ \int dx \int dy  \int dz \psi(x,y,z)  v(x,y,z) \psi(x,y,z) $$
and also for matrix elements between different wavefunctions
$$ \int dx \int dy  \int dz \psi_i(x,y,z)  v(x,y,z) \psi_j(x,y,z) $$
in case they are needed. Because of this remarkable achievement of the filter $\omega$
we call it the magic filter.

Following the same guidelines as the kinetic energy filters, the smoothed real space values $\tilde{\psi}_{i,j,k}$ of a wavefunction $\psi$ are calculated by performing a product of three one-dimensional convolutions with the magic filters along the $x$, $y$ and $z$ directions.
For the scaling function part of the wavefunction the corresponding formula is
$$\tilde{\psi}_{i_1,i_2,i_3} = 
   \sum_{j_1,j_2,j_3} s_{j_1,j_2,j_3}  v_{i_1-2 j_1}^{(1)} v^{(1)}_{i_2-2 j_2} v^{(1)}_{i_3-2 j_3} $$
where $v_{i}^{(1)}$ is the filter that maps a scaling function on a double resolution
grid. Similar convolutions are needed for the wavelet part.
The calculation is thus similar to the treatment of the Laplacian in the kinetic energy.

Once we have calculated $\tilde{\psi}_{i,j,k}$ the approximate expectation value $\epsilon_V$ of the local potential $v$ for a wavefunction $\psi$ is obtained by simple summation on the double resolution real space grid:
$$ \epsilon_v = \sum_{j_1,j_2,j_3} 
                \tilde{\psi}_{j_1,j_2,j_3}  v_{j_1,j_2,j_3} \tilde{\psi}_{j_1,j_2,j_3} $$

\subsection{Treatment of the non-local pseudopotential}\label{nonlocalpseudosection}
The energy contributions from the non-local pseudopotential have for each angular
moment $l$ the form
$$\sum_{i,j} \langle \psi | p_i \rangle h_{ij} \langle p_j | \psi \rangle $$
where $| p_i \rangle$ is a pseudopotential projector.
Once applying the Hamiltonian operator, the application of one projector on the wavefunctions  requires the calculation of
$$  |\psi\rangle \rightarrow |\psi \rangle + \sum_{i,j} |p_i \rangle h_{ij}  \langle p_j | \psi \rangle\;. $$
If we use for the projectors the representation of Eq.~(\ref{eq:daub.2}) (i.e., the same as for
the wavefunctions) both operations are trivial to perform. Because of the orthogonality of the basis set we just have to calculate scalar products among the coefficient vectors and to update the wavefunctions.
The scaling function and wavelet expansion coefficients for
the projectors are given by~\cite{G98}
\begin{equation}
 \int p(\mathbf r)\, \phi_{i_1,i_2,i_3}({\bf r}) \rm d{\bf r}\;, \, \, \, 
 \int p(\mathbf r)\, \varphi^{\nu}_{i_1,i_2,i_3}({\bf r}) \rm d{\bf r}\;.
\end{equation}

The GTH-HGH pseudopotentials~\cite{GTH96,HGH98}  
have projectors which are written in terms of gaussians times polynomials. This form of projectors is particularly convenient to be expanded in the Daubechies basis.
In other terms, since the general form of the projector is
$$\langle \mathbf r |p \rangle = e^{- c r^2}  x^{\ell_x}y^{\ell_y}z^{\ell_z}\;, $$
the 3-dimensional integrals can be calculated easily since they can be factorized into a product of 3 one-dimensional integrals.
\begin{eqnarray}
   \int \langle \mathbf r |p \rangle  \phi_{i_1,i_2,i_3}({\bf r}) \rm d{\bf r} &=&
  W_{i_1}(c,\ell_x) W_{i_2}(c,\ell_y) W_{i_3}(c,\ell_x)\;,\\
   W_{j}(c,\ell) &=& \int_{-\infty}^{+\infty} e^{- c t^2} t^\ell \phi(t/h-j) \rm d t
\end{eqnarray}

The one-dimensional integrals are calculated in the following way. We first calculate the
scaling function expansion coefficients for scaling functions on a one-dimensional grid that is 16 times
denser. The integration on this dense grid is done by
the well-known quadrature introduced in \cite{JMNK99}, that coincides with the magic filter 
\cite{NG06}. This integration scheme based on the magic filter has a convergence rate of 
$h^{16}$ and we gain therefore a factor of $16^{16}$ in accuracy by going to a denser grid. 
This means that the expansion coefficients are for reasonable grid spacings $h$ accurate to 
machine precision. After having obtained the expansion coefficients with respect to the fine 
scaling functions we obtain the expansion coefficients with respect to the scaling functions 
and wavelets on the required resolution level by one-dimensional fast wavelet transformations.
No accuracy is lost in the wavelet transforms and our representation
of the projectors is therefore typically accurate to nearly machine
precision. In order to treat with the same advantages other pseudopotentials which are not 
given under the form of gaussians it would be necessary to approximate them by a small number 
of gaussians.

\subsection{The Poisson operator}
\label{section:poisson}
Solving the Poisson equation for an arbitrary charge distribution
is a non-trivial task, and is of major importance in many field of science,
especially in the field of computational chemistry.  A huge effort has been
put into making efficient Poisson solvers, and usual real-space approaches
includes finite difference (FD) and finite element (FE) methods.  FD is a 
grid-based method, which is solving the equations iteratively on a
discrete grid of pointvalues, while FE is expanding the solution in a basis
set, usually by dividing space into cubic cells and allocate a polynomial
basis to each cell.

It is well-known fact that the electronic density in molecular systems is
rapidly varying in the vicinity of the atomic nuclei, and a usual problem
with real-space methods is that an accurate treatment of the system requires
high resolution of gridpoints (FD) or cells (FE) in the nuclear regions.
Keeping this high resolution uniformly throughout space would yield unnecessary
high accuracy in the interatomic regions, and the solution of the Poisson
equation for molecular systems is demanding a {\em multiresolution} framework
in order to achieve numerical efficiency.
This chapter is
concerned with a way of doing DFT and TD-DFT calculations, one where the
multiresolution character is inherent in the theory, namely using
wavelet bases.

In order to evaluate the Hartree potential, we need to rewrite the
standard Poisson equation to an integral form. The equation, in
its differential form, is given as
\begin{equation}
  \nabla^2 v({\bf{x}}) = 4 \pi \rho({\bf{x}}) \, ,
  \label{eq:ps.1}
\end{equation}
where $\rho({\bf{x}})$ is the known (charge) distribution, and
$v({\bf{x}})$ is the unknown (electrostatic) potential.  It is a
standard textbook procedure to show that the solution can be written as the
integral
\begin{equation}
  v({\bf{x}}) = \int G({\bf{x,y}}) \rho({\bf{y}}) d{\bf{y}} \, ,
  \label{eq:ps.2}
\end{equation}
where $G({\bf{x,y}})$ is the Green's function which is the solution to the
{\em fundamental} equation with {\em homogeneous} (Dirichlet) boundary
conditions
\begin{eqnarray}
  \nabla^2 G({\bf{x,y}}) & =  & \delta({\bf{x-y}}) \, \nonumber \\
  G({\bf{x,y}}) & = & 0 \, , \mbox{} {\bf{x}} \in  \mbox{boundary}
  \label{eq:ps.3}
\end{eqnarray}
This equation can be solved analytically and the Green's function is
given (in three dimensions) simply as
\begin{equation}
  G({\bf{x,y}}) = \frac{1}{||{\bf{x,y}}||} \, ,
  \label{eq:ps.4}
\end{equation}
This is the  well-known potential arising from a point charge located
in the position ${\bf{y}}$, which is exactly what Eq.~(\ref{eq:ps.3}) describes.

\subsection{Numerical separation of the kernel}
\label{sec:kernel}
The Green's function kernel as it is given in Eq.~(\ref{eq:ps.4}) is not
separable in the cartesian coordinates.  However, since we are working with
finite precision we can get by with an {\em approximate} kernel as long as
the error introduced with this approximation is less than our overall
accuracy criterion.  If we are able to obtain such a {\em numerical} 
separation of the kernel, the operator can be applied
in one direction at the time, allowing us to use the expressions 
derived above for one-dimentional integral operators to solve the
three-dimensional Poisson equation. This is of great importance because it
reduces the scaling behavior to become linear in the dimension of the system.

The Poisson kernel can be made separable by expanding it  as a sum of Gaussian
functions, specifically
\begin{equation}
  \frac{1}{r} \simeq \sum_{k=1}^{M_\epsilon} \omega_k e^{-p_k r^2}\;.
  \label{eq:ps.5}
\end{equation}
where $\omega_k$ and $p_k$ are parameters that needs to be determined, 
and the number of terms $M_\epsilon$, called the separation rank,
depends on the accuracy requirement and on what interval this
expression needs to be valid.  Details of how to obtain this
expression can be found in \cite{GDNGB06,GDG07}, and will not be treated here,
but it should be mentioned that the separation rank is usually
in the order of 100, e.g, it requires $M_\epsilon = 89$ to reproduce
$\frac{1}{r}$ on the interval  $[10^{-9}\,, 1 ]$ in three dimensions with
error less than $\epsilon = 10^{-8}$.

\begin{figure}[h!]
\begin{center}
\includegraphics[angle=0,width=0.8\textwidth]{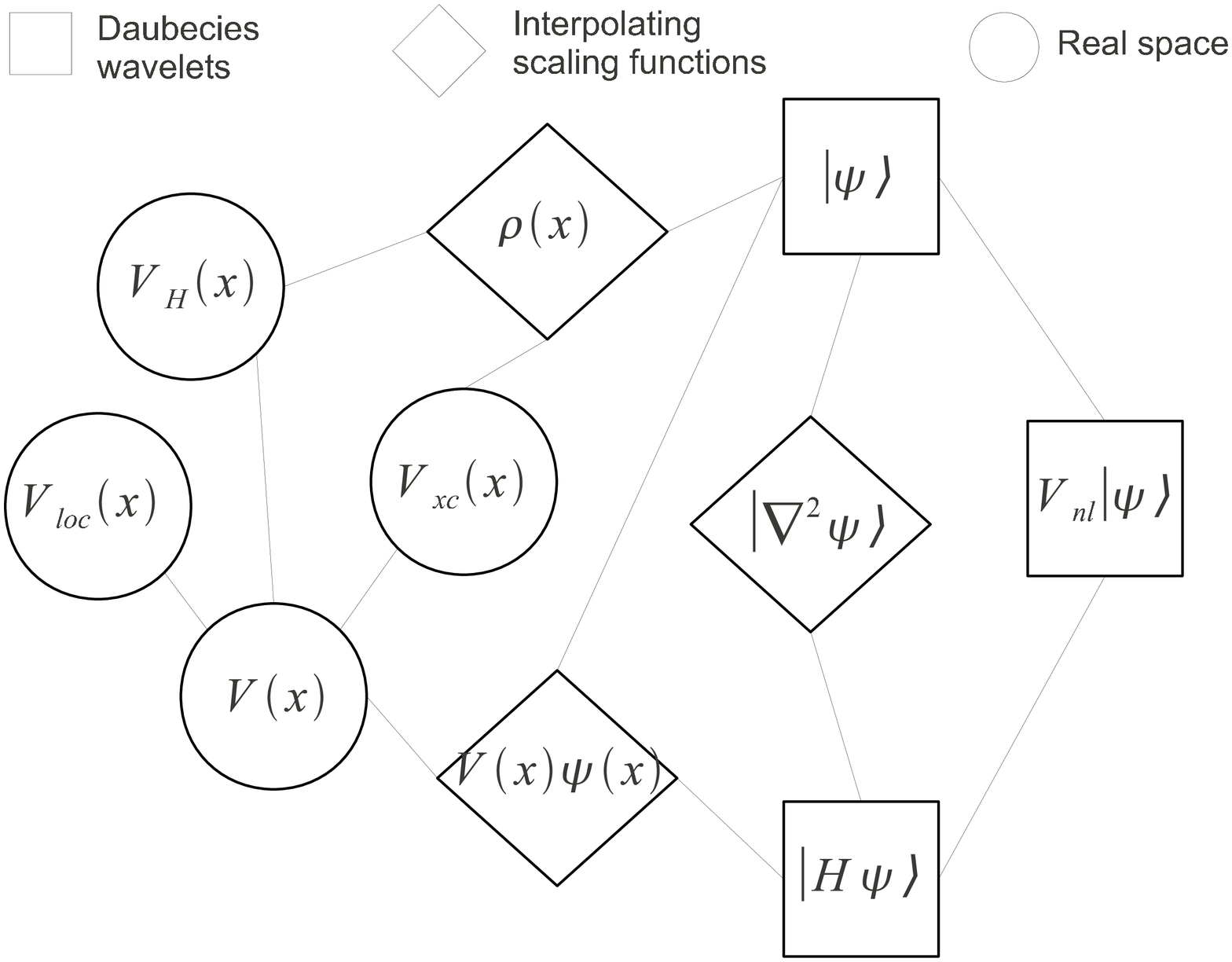}
\end{center}
\caption{ Operations performed in {\sc BigDFT}
          \label{fig:bigdft-operations}
}
\end{figure}

Finally, figure ~{\ref{fig:bigdft-operations}} summarize this complete section into a flow-chart type diagram. 
This kind of explanation is necessary for begineers because there are  different functions used for the different operations in {\sc BigDFT}. As one can see
from the figure, The KS wavefunctions $|\psi\rangle$ are expressed in terms of Daubechies
wavelets and the projection of Hamiltonian $V_{nl}|\psi\rangle$ and of pseudopotential  
operators $|H\psi\rangle$  also expressed using Daubechies wavelets. The rest of the
operations such as kinetic energy $|\nabla^2\psi\rangle$,  potential energy operator $V(x)\psi(x)$, and 
the local densities $\rho(x)$ are all expressed using interpolating scaling functions, in which
the Hartree $V_H(x)$, local part potential energy $V_{loc}(x)$ and xc operations $V_{xc}(x)$ were performed in real space.     
The interconnecting lines between different operations represents the
transformation between Daubechies wavelets-to-ISFs or the transformation of
real space-to-fourier space representation. 
\section{{\sc BigDFT} and TD-DFT}
\label{sec:bigdft-tddft}
We want to solve Casida's equation \cite{C95},
\begin{equation}
\label{eq:casida-c}
  \left[ \left ( \begin{array}{cc} {\bf A}({\cal{\omega}}) &{\bf  B}({\cal{\omega}}) \\ {\bf B^*}({\cal{\omega}}) & {\bf A^*}({\cal{\omega}}) \end{array} \right ) 
  - \omega \left ( \begin{array}{cc} 1 & 0 \\ 0 & -1 \end{array} \right ) \right ] 
   \left ( \begin{array}{c} {\bf X} \\ {\bf Y} \end{array}\right ) = 0  \, ,
\end{equation}
where $ {\bf X}$ and ${\bf Y} $ represents the pseudo eigenvectors;
the matrices ${\bf A}$ and ${\bf B}$ are defined as
\begin{equation}
\label{eq:casida-a}
  {\bf A}_{ai\sigma,bj\tau} = \delta_{ab} \delta_{ij}\delta_{\sigma\tau} (\epsilon_a -\epsilon_i) 
   + {\bf K}_{ai\sigma,bj\tau}({\cal{\omega}}) \, ,
\end{equation}
and,
\begin{equation}
\label{eq:casida-b}
  {\bf B}_{ai\sigma,bj\tau} = {\bf K}_{ai\sigma,jb\tau}({\cal{\omega}}) \, ,
\end{equation}
in which the integral form of the coupling matrix $\bf K$ is given by,
\begin{eqnarray}
  & & {\bf K}_{pq\sigma,rs\tau}   =   
  \int \int \Psi_{p\sigma}^{*}(\vec r) \Psi_{q\sigma}(\vec r)  \nonumber \\
  & & \left [ \frac{1}{|\vec r - \vec r'|} + \frac {\partial^2 E_{xc}[\rho]}{\partial \rho_\sigma(\vec r)\partial \rho_\tau (\vec r')} \right ] 
  \Psi_{r\tau}(\vec r')\Psi_{s\tau}^{*}(\vec r') d\vec r d\vec r' \, .
  \label{eq:k-int}
\end{eqnarray}
The universal adiabatic approximation is applied to Eq.~(\ref{eq:k-int}) to remove the frequency dependence of the kernel.

The electronic transitions occur with an infinitesimal perturbation obtains
the above described non-Hermitian eigenvalue Eq.~(\ref{eq:casida-c}).
Where the response is due to a real spin independent external
perturbation, and the actual response is described as the real density response.
However, an unitary transformation is necessary to convert
Eq.~(\ref{eq:casida-c}) into the real eiganvalue problem.
In Eq.~(\ref{eq:casida-c}), all occupied-occupied and virtual-virtual element
contributions are zero whereas only the elements that are from
virtual-occupied and occupied-virtual parts are taken into account.
Moreover if we only restricted to virtual-occupied elements
and neglecting the occupied-virtual elements of Eq.~(\ref{eq:casida-c})
leads to a Hermitian eigenvalue equation of the dimension one-half of
that TD-DFT working equation is said to be Tamm-Dancoff approximation (TDA)  
and it is written as,
\begin{equation}
\label{eq:tda}
   {\bf AX} = \omega {\bf X} \, ,
\end{equation}
where ${\bf A}$ is as same as in Eq.~(\ref{eq:casida}).
The matrix {\bf A} is just restricted to number of single excitations.

However, the explicit form of Eq.~(\ref{eq:tda}) is,
\begin{equation}
\Omega(\omega)\vec F_I = \omega^2\vec F_I \, ,
\label{eq:tda-imp}
\end{equation}
where
\begin{eqnarray}
\label{eq:imp}
\Omega_{ia\sigma,jb\tau} & = & \delta_{ia}\delta_{jb}\delta_{\sigma\tau}{(\epsilon_{a\sigma} -\epsilon_{i\sigma})^2} + \\ \nonumber 
&   &2 \sqrt{(\epsilon_{a\sigma} -\epsilon_{i\sigma})} K_{ia\sigma,jb\tau} \sqrt{(\epsilon_{a\sigma} -\epsilon_{i\sigma})} \, ,
\end{eqnarray}
where $\epsilon_{i\sigma} -\epsilon_{a\sigma}$
is the energy eigenvalue differences of ${\it i}^{th}$ and
${\it a}^{th}$ states.
Solving Eqs.~(\ref{eq:tda-imp})  yields TD-DFT
 excitation energies $\omega$ and $\vec F_I$'s are the corresponding oscillator strengths which are defined from the transition dipole moments.

\subsection{Calculation of Coupling Matrix}
We are now in a position to understand the construction of the coupling matrix Eq.~(\ref{eq:k-int}) in our
implementation of TD-DFT in {\sc BigDFT}, which we split into the Hartree andxc parts,
\begin{equation}
  K_{ai\sigma,bj\tau} = K^H_{ai\sigma,bj\tau} + K^{xc}_{aj\sigma,bj\tau} \, .
  \label{eq:BigDFT.6}
\end{equation}
Instead of calculating the Hartree part of coupling matrix directly as,
\begin{equation}
  K^H_{ai\sigma,bj\tau}  = \int \int \psi_{a\sigma}^{*}({\bf r}) \psi_{i\sigma}({\bf r})
         \frac{1}{\vert {\bf r}  - {\bf r}' \vert}
         \psi_{b\tau}({\bf r'}) \psi_{j\tau}^{*}({\bf r'}) \, d{\bf r} d{\bf r}' \, ,
  \label{eq:BigDFT.7}
\end{equation}
we express the coupling matrix element as,
\begin{equation}
  K^H_{ai\sigma,bj\tau}  = \int \psi_{a\sigma}^{*}({\bf r}) \psi_{i\sigma}({\bf r})
                v_{bj\tau}({\bf r}) \, d{\bf r} \, ,
  \label{eq:BigDFT.8}
\end{equation}
where,
\begin{equation}
  v_{ai\sigma}({\bf r}) = \int \frac{\rho_{ai\sigma}({\bf r})}{\vert {\bf r} - {\bf r}'|} \, d{\bf r}' \, ,
  \label{eq:BigDFT.9}
\end{equation}
and,
\begin{equation}
  \rho_{ai\sigma}({\bf r}) = \psi_{a\sigma}^{*}({\bf r})\psi_{i\sigma}({\bf r}) \, .
  \label{eq:BigDFT.10}
\end{equation}
The advantage of doing this is that, although $\rho_{ai\sigma}$ and $v_{ai\sigma}$
are neither real physical charge densities nor real physical potentials, they still
satisfy the Poisson equation,
\begin{equation}
  \nabla^2 v_{ai\sigma}({\bf r}) = - 4 \pi \rho_{ai\sigma}({\bf r}) \, ,
  \label{eq:BigDFT.11}
\end{equation}
and we can make use of whichever of the efficient wavelet-based Poisson solvers
already available in {\sc BigDFT}, is appropriate for the boundary conditions of our
physical problem.

Once the solution of Poisson's equation, $v_{ai\sigma}({\bf r})$, is known, we can then calculate the
Hartree part of the kernel according to Eq.~(\ref{eq:BigDFT.8}).  Inclusion of the xc kernel is accomplished by evaluating,
\begin{equation}
  K_{ai\sigma,bj\tau} = \int M_{ai\sigma}({\bf r}) \rho_{bj\tau}({\bf r}) \, d{\bf r}
  \, ,
  \label{eq:BigDFT.12}
\end{equation}
where,
\begin{equation}
  M_{ai\sigma}({\bf r}) = v_{ai\sigma}({\bf r}) 
   +  \int \rho_{ai\sigma} ({\bf r}')  f_{xc}^{\sigma,\tau}({\bf r},{\bf r}')  \, d{\bf r}'
  \, .
  \label{eq:BigDFT.13}
\end{equation}
We note that $f_{xc}^{\sigma,\tau}({\bf r},{\bf r}') = f_{xc}^{\sigma,\tau}({\bf r},{\bf r}') \delta({\bf r}-{\bf r}')$ for the LDA, so that no integral need actually be carried out in evaluating $M_{ai\sigma}({\bf r})$.
The integral in Eq.~(\ref{eq:BigDFT.12}) is, of course, carried out
numerically in practice as a discrete summation.

\section{Results}
\label{sec:results}

We now wish to illustrate a bit how wavelet calculations work in the {\sc BigDFT} program.
Comparison will be made against results obtained with the GTO-based program {\sc deMon2k}.
This work is very similar to our previous work reporting the first implementation of wavelet-based
TD-DFT with illustration for N$_2$ and application to the absorption spectrum of a medium-sized 
organic molecule of potential biomedical use as a fluorescent probe~\cite{NGC+11}.  Here however
we will present new {\sc BigDFT} results for a different small molecule, namely carbon monoxide.
Though CO is roughly isoelectronic with N$_2$, CO has the interesting feature of having a 
low-lying bright state in its absorption spectrum.

\subsection{Computational Details}

Calculations were carried out with {\sc deMon2k} and {\sc BigDFT} with the LDA-optimized bond length of 1.129 {\AA}.

\subsubsection{\sc deMon2k}

{\sc deMon2k} resembles a typical GTO-based quantum chemistry program in that all the integrals other than the
xc-integrals, can be evaluated analytically. In particular, {\sc deMon2k} has the important advantage that it accepts the
popular GTO basis sets common in quantum chemistry and so can benefit from the experience in basis set construction
of a large community built up over the past 50 years or so. In the following, we have chosen to use the well-known
correlation-consistent basis sets for this study \cite{F96,SDE+07}. (Note, however, that the correlation-consistent 
basis sets used in {\sc deMon2k} lack f and g functions but are otherwise exactly the same as the usual ones.) 
The advantage of using these particular basis sets is that there is a clear hierarchy as to quality.

An exception to the rule that integrals are evaluated analytically in {\sc deMon2k} are the xc-integrals (for the xc-energy,
xc-potential, and xc-kernel) which are evaluated numerically over a Becke atom-centered grid. This is important because 
the relative simplicity of evaluating integrals over a grid has allowed the rapid implemenation of new functionals
as they were introduced. We made use of the fine fixed grid in our calculations.

As described so far, {\sc deMon2k} should have ${\cal O}(N^4)$ scaling because of the need to evaluate 4-center integrals.
Instead {\sc deMon2k} uses a second atom-centered auxiliary GTO basis to expand the charge density. This allows the
the elimination of all 4-center integrals so that only 3-center integrals remain for a formal ${\cal O}(N^3)$ scaling. 
In practice, integral prescreening leads to ${\cal O}(N M)$ scaling where M is typically between 2 and 3. We made use of 
the A3 auxiliary basis set from the {\sc deMon2k} automated auxiliary basis set library.

All calculations were performed using standard {\sc deMon2k} default criteria. 
The implementation of TD-DFT in {\sc deMon2k} is described in Ref.~\cite{IFP+06}.
(The charge density conservation constraint is no longer used in {\sc deMon2k} TD-DFT calculations.)
Although full TD-LDA calculations are possible with {\sc deMon2k}, the TD-LDA calculations reported here all made 
use of the TDA.

\subsubsection{\sc BigDFT}

\begin{figure}[ht]
\centering
\subfigure[H$_2$O in a box]{
\includegraphics[angle=0,width=0.3\textwidth]{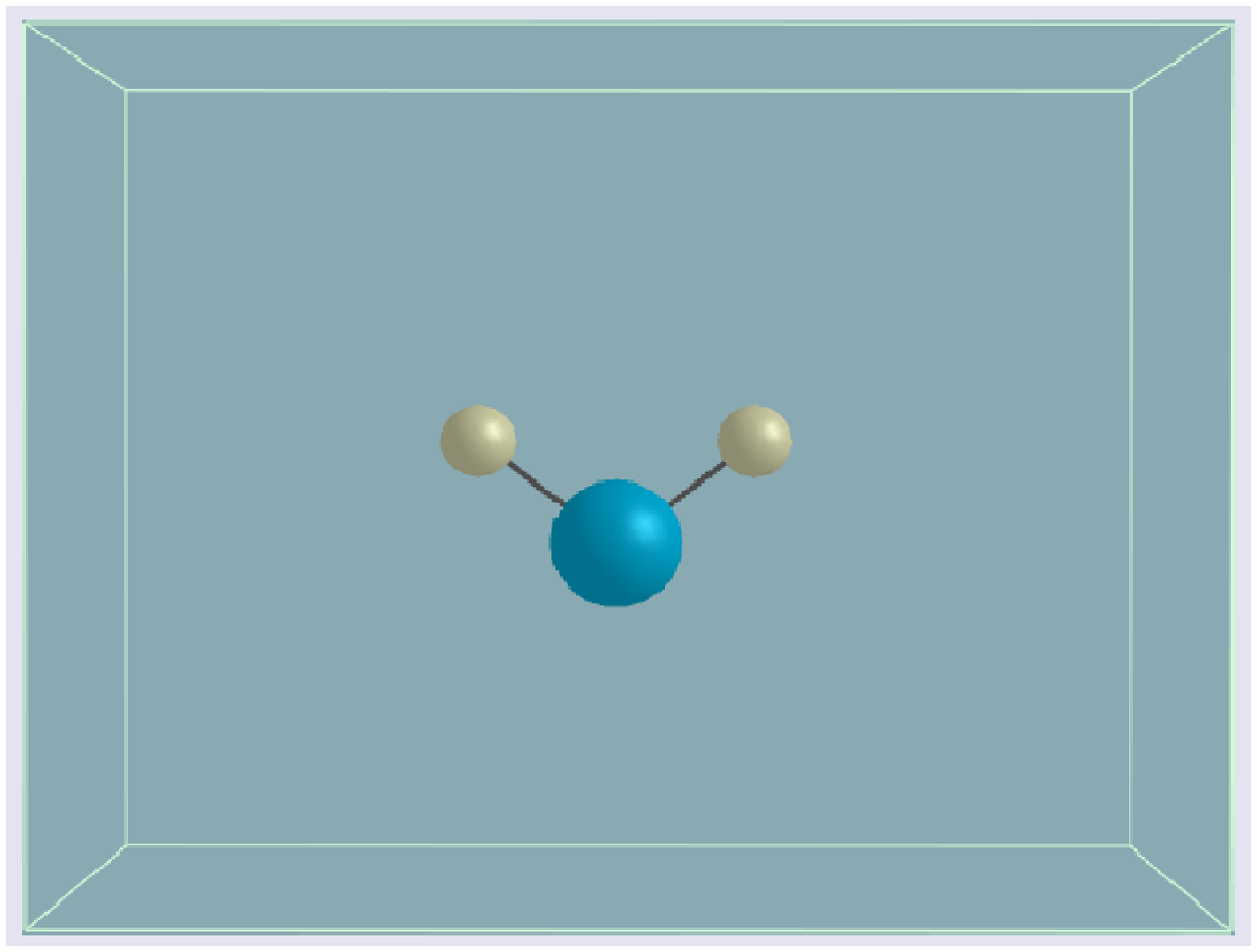}
\label{fig:H$_2$O}
}
\subfigure[H$_2$O in a box showing fine grid resolution]{
\includegraphics[angle=0,width=0.3\textwidth]{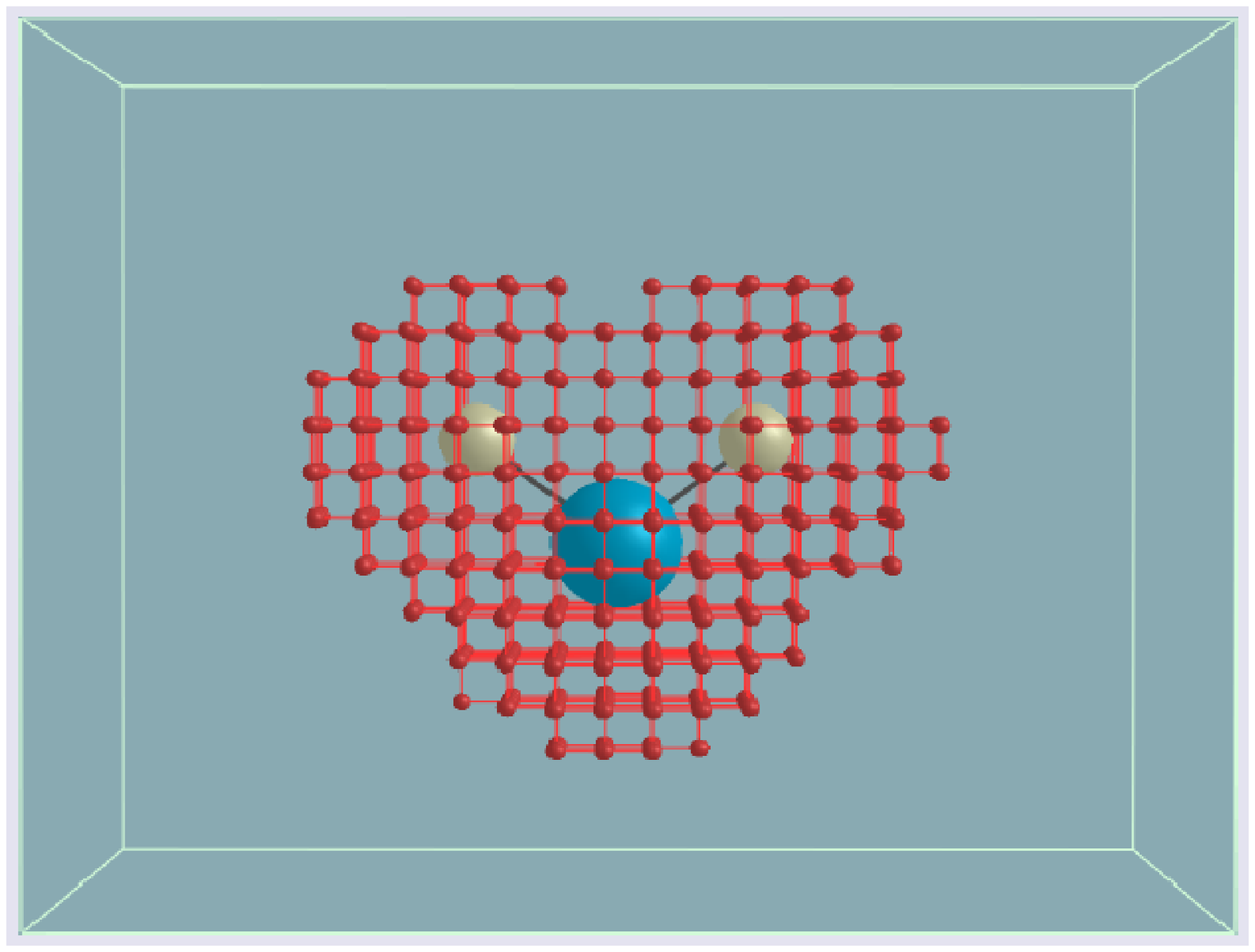}
\label{fig:H$_2$O-fine-grid}
}
\subfigure[H$_2$O in a box showing coarse grid resolution]{
\includegraphics[angle=0,width=0.3\textwidth]{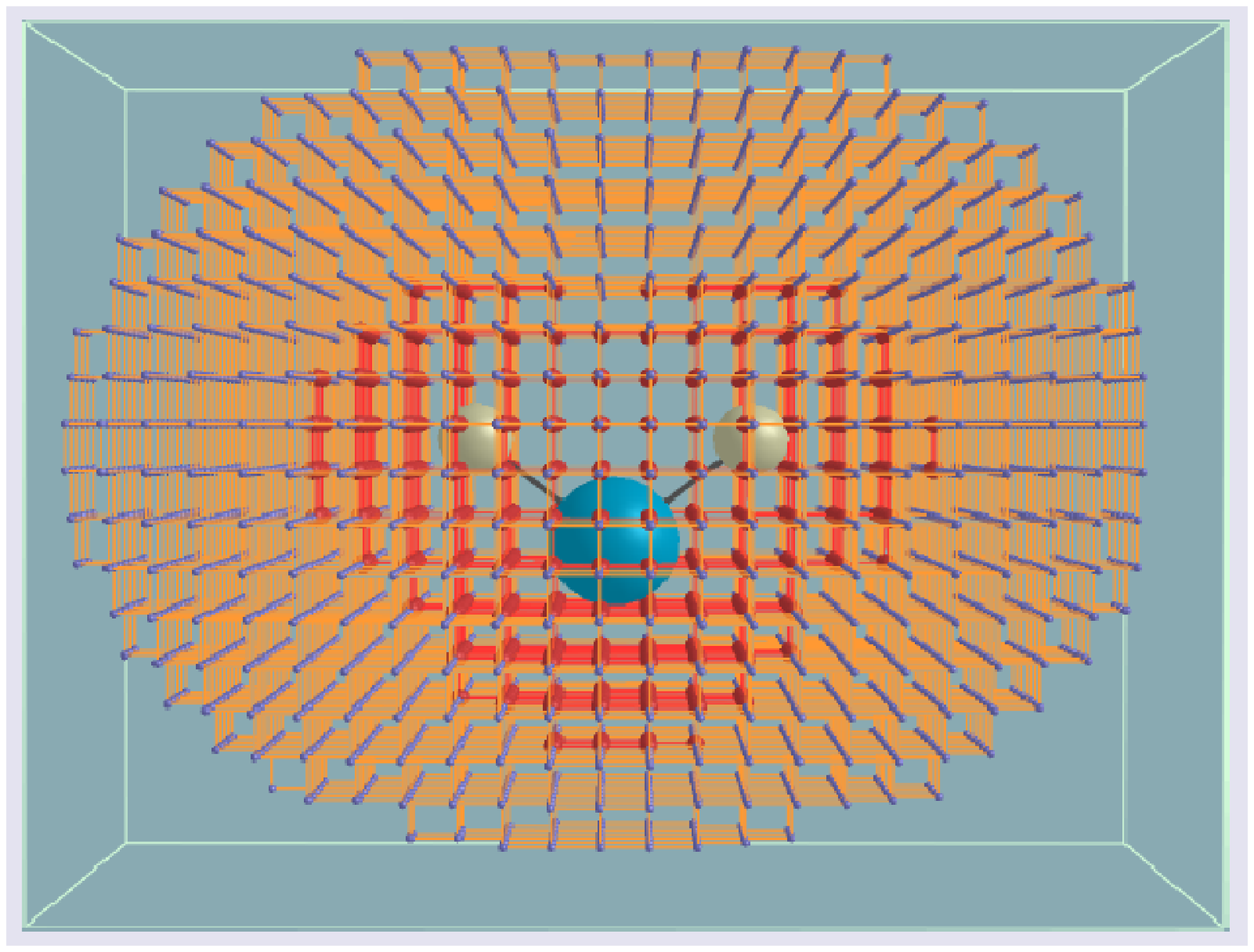}
\label{fig:H$_2$O-coarse-grid}
}
\label{fig:Grid-in-bigdft}
\caption[Adaptive grid in {\sc BigDFT}]{Adaptive grid in {\sc BigDFT} \subref{fig:H$_2$O}, \subref{fig:H$_2$O-fine-grid} and \subref{fig:H$_2$O-coarse-grid}}
\end{figure}
 

The main thing to vary in {\sc BigDFT} is the grid which is of more profound importance
than in {\sc deMon2k} because it is the grid which supports the wavelets.
Figures~\ref{fig:H$_2$O}, \ref{fig:H$_2$O-fine-grid}, and \ref{fig:H$_2$O-coarse-grid}
give an idea of what the grid looks like for the small familiar molecule of water.  Conceptually
the molecule is in a very large box (Fig.~\ref{fig:H$_2$O}.)  A fine grid is placed in the regions
of high electron density around the molecule (Fig.~\ref{fig:H$_2$O-fine-grid}.)  A coarse grid is
used in a larger region where the electron density varies more slowly (Fig.~\ref{fig:H$_2$O-coarse-grid}.)
The {\sc BigDFT} grid is characterized by the triple $h_g$/crmult/frmult.  The first number in the triple
($h_g$) is a real number which specifies the nodes of the grid in atomic units.  The second number 
(the integer-valued crmult) is the coarse grid multiplier.  And the third number (the integer-valued 
frmult) is the fine grid multiplier.  Two points must be clearly understood when looking at 
Figs.~\ref{fig:H$_2$O}, \ref{fig:H$_2$O-fine-grid}, and \ref{fig:H$_2$O-coarse-grid}.  The first is that, 
while the box may determine the limits of the grid, the grid does not have the shape of the box and there are
no basis functions where there are no grid points.  This means that we are not dealing with box boundary 
conditions, but rather with effective boundary conditions which reflect the shape of the molecule.
The other point which is not brought out by our explanation is that the {\sc BigDFT} grid is adaptive
in the sense that additional fine grid points are added during the calculation as they are needed
to maintain and improve numerical precision.  

The implementation of TD-DFT in {\sc BigDFT} is described in Ref.~\cite{NGC+11}.

\subsection{Orbital Energies}

Possibly the most remarkable property of wavelets is how rapidly they converge to the basis set limit.
Let us illustrate this by comparing highest-occupied molecular orbital (HOMO) and  
lowest-unoccupied molecular orbital (LUMO) energies calculated with {\sc deMon2k} and {\sc BigDFT}.
The difference of these two energies is the HOMO-LUMO gap, $\Delta \epsilon_{HOMO-LUMO}$.

\begin{table}
\small
  \caption{Basis set dependence of the HOMO and LUMO energies and of the HOMO-LUMO gap (eV) calculated
           using {\sc deMon2k}.  \label{tab:deMon-basis-set-dependence}}
  \begin{center}
  \begin{tabular}{cccc}
  \hline \hline
  Basis Set & $-\epsilon_{HOMO}$ & $-\epsilon_{LUMO}$
  &  $\Delta \epsilon_{HOMO-LUMO}$  \\
  \hline
  STO-3G         & -5.5350   &  1.2428     & 4.2922     \\
  \multicolumn{4}{c}{ } \\
  DZVP           &  -8.9271 &  -2.0942     & 6.8329    \\
  TZVP           & -9.0287  &  -2.1902     & 6.8385    \\
  \multicolumn{4}{c}{ } \\
  CC-PVDZ        & -8.6729  &  -1.7823     & 6.8906    \\
  CC-PVTZ        &  -9.0419 &  -2.1195    &  6.9224   \\
  CC-PVQZ        &  -9.0944 &  -2.1971     & 6.8973   \\
  CC-PV5Z        &  -9.1169 &  -2.2400     & 6.8769    \\
 \multicolumn{4}{c}{ } \\
  CC-PCVDZ       &  -8.6905 &  -1.7922     & 6.8983    \\
  CC-PCVQZ       &  -9.0957 &  -2.1988     & 6.8969    \\
  CC-PCVTZ       &  -9.0371 &  -2.1165     & 6.9206    \\
  CC-PCV5Z       & -9.1172  &  -2.2401     & 6.8771    \\
  \multicolumn{4}{c}{ } \\
  AUG-CC-PVDZ    &  -9.0910 &  -2.2345     & 6.8565    \\
  AUG-CC-PVQZ    &  -9.1286 &  -2.2567     & 6.8719   \\
  AUG-CC-PVTZ    &  -9.1306 &  -2.2535     & 6.8771    \\
  AUG-CC-PV5Z    &  -9.1289 &  -2.2606     & 6.8683    \\
  \multicolumn{4}{c}{ } \\
  AUG-CC-PCVDZ   &  -9.0987 &  -2.2371     &  6.8616   \\
  AUG-CC-PCVTZ   &  -9.1316 &  -2.2554     &  6.5776   \\
  AUG-CC-PCVQZ   & -9.1293  & -2.2574      &  6.8719   \\
  AUG-CC-PCV5Z   & -9.1291  & -2.2607      &  6.8684   \\
  \hline \hline
  \end{tabular}
  \end{center}
\end{table}
\begin{table} 
\small
  \caption{Basis set dependence of the HOMO and LUMO energies and of the HOMO-LUMO gap (eV) calculated
           using {\sc BigDFT}.  \label{tab:BigDFT-basis-set-dependence}}
\begin{center}
  \begin{tabular}{cccc}
  \hline
  h$_g$$^a$/m$^b$/n$^c$
  & $-\epsilon_{HOMO}$ & $-\epsilon_{LUMO}$
  &  $\Delta \epsilon_{HOMO-LUMO}$  \\
  \hline
  0.4/6/8  &  -9.0976  & -2.1946  & 6.9029 \\
  0.4/7/8  &  -9.1014 & -2.2028  & 6.8985 \\
  0.4/8/8  &  -9.1017 & -2.2044  & 6.8971 \\
  0.4/9/8  & -9.1017   & -2.2049  & 6.8967 \\
  0.4/10/8 & -9.1017   & -2.2049  & 6.8966 \\
  0.3/7/8  &  -9.1022  & -2.2056  & 6.8964 \\
  0.3/8/8  &   -9.1025 &  -2.2073 & 6.8950 \\
  \hline
  \end{tabular}
\end{center}
$^a$Grid spacing of the cartesian grid in atomic units.\\
$^b$Coarse grid multiplier (crmult).\\
$^c$ Fine grid multiplier (frmult).
\end{table}
Consider first how {\sc deMon2k} calculations of $\Delta \epsilon_{HOMO-LUMO}$,
evolve as the basis set is improved (Table~\ref{tab:deMon-basis-set-dependence}.)
Convergence to the true HOMO-LUMO LDA gap is expected with systematic improvement within the series:
(i) double zeta plus valence polarization (DZVP) $\rightarrow$ triple zeta plus valence polarization (TZVP),
(ii) augmented correlation-consistent double zeta plus polarization plus diffuse on all
atoms (AUG-CC-PCVDZ) $\rightarrow$ AUG-CC-PCVTZ (triple zeta) $\rightarrow$ AUG-CC-PCVQZ (quadruple zeta)
$\rightarrow$ AUG-CC- PCV5Z (quintuple zeta),
(iii) augmented correlation-consistent valence double zeta plus polarization plus diffuse
(AUG-CC-PVDZ) $\rightarrow$ AUG-CC-PVTZ $\rightarrow$ AUG-CC-PVQZ $\rightarrow$ AUG-CC-PV5Z,
(iv) correlation-consistent double zeta plus polarization plus tight core (CC-PCVDZ) $\rightarrow$ CC-PCVTZ
$\rightarrow$ CC-PCVQZ $\rightarrow$ CC-PCV5Z, and (v) correlation-consistent valence double zeta plus
polarization on all atoms (CC-PVDZ) $\rightarrow$ CC-PVTZ $\rightarrow$ CC-PVQZ $\rightarrow$ CC-PV5Z.
There is a clear tendency in the correlation-consistent basis sets to tend towards values of -9.13 eV for the HOMO
energy, -2.26 eV for the LUMO energy, and 6.87 eV for $\Delta \epsilon_{HOMO-LUMO}$, with adequate convergance
achieved with the AUG-CC-PVQZ basis set.

Now let us turn to BigDFT (Table~\ref{tab:BigDFT-basis-set-dependence}). Calculations were done for several different 
grids, including the high-resolution combination 0.3/8/8 and the low-resolution combination of 0.4/6/8.
Remarkably, except for the very lowest quality grid 0.4/6/8, there is essentially no difference between results 
obtained with the two grids (and even the 0.4/6/8 grid gives nearly converged results.)  The results are also
quite close to, but not identical to those obtained with the {\sc deMon2k} program.  The reason for the small
differences between the converged results obtained with the two programs is more difficult to trace as it might 
be due to the auxiliary basis approximation in {\sc deMon2k} or to the use of pseudopotentials in {\sc BigDFT} 
or perhaps to still other program differences. The important point is that differences are remarkably small.

\subsection{Excitation Energies}
Orbital energy differences provide a first estimate for excitation energies.  In this case, we would
expect to see the HOMO $\rightarrow$ LUMO excitation at $\Delta \epsilon_{HOMO-LUMO} \approx 6.9$ eV (6.87 eV for {\sc deMon2k}
and 6.90 eV for {\sc BigDFT}.)  A better estimate is provided by the two-orbital two-electron model (TOTEM) 
\cite{C95,CGG+00,C09,CH12} for the singlet (S) and triplet (T) transition from orbital $i$ to orbital $a$,
\begin{eqnarray}
  \hbar \omega_{i \rightarrow a}^T & = & \Delta \epsilon_{i \rightarrow a} + (ia \vert f_{xc}^{\alpha,\alpha} 
              - f_{xc}^{\alpha,\beta} \vert ai ) \nonumber \\
  \hbar \omega_{i \rightarrow a}^S & = & \Delta \epsilon_{i \rightarrow a} + (ia \vert 2 f_H + f_{xc}^{\alpha,\alpha}
              + f_{xc}^{\alpha,\beta} \vert ai ) \, ,
  \label{eq:results.1}
\end{eqnarray}
where
\begin{equation}
  \Delta \epsilon_{i \rightarrow a} = \epsilon_a - \epsilon_i \, .
  \label{eq:results.2}
\end{equation}

\begin{table} 
\small
  \caption{Comparison of  lowest excitation energies of CO (in eV) calculated using {\sc BigDFT} and {\sc deMon2k}
            and with experiment. \label{tab:co-excitation-energy}}
\begin{center}
  \begin{tabular}{lccc}
  \hline \hline
  State             & {\sc BigDFT}$^a$
                    &  {\sc deMon2k}$^b$
                    & Experiment$^c$\\
  \hline
  1$^3\Sigma^{-}$  & 9.84          & 9.85                  & 9.88 \\
  1$^3\Delta$      & 9.17          & 9.21                  & 9.36\\
  1$^1\Pi$         & 8.94          & 8.42                  & 8.51\\
  1$^3\Sigma^{+}$  & 8.94          & 8.54                  & 8.51 \\
  1$^3\Pi$         & 6.47          & 6.05                 & 6.32 \\
  \hline \hline
  \end{tabular}
\end{center}
  $^a$ Present work (TD-LDA/TDA) using AUG-CC-PCQZ basis set. \\
  $^b$ Present work (TD-LDA/TDA) using 0.3/8/8 grid.  \\
  $^c$ Taken from Ref.~\cite{CS00}.
\end{table}
The TOTEM model often works surprisingly well for small molecules because, unlike the Hartree-Fock approximation which
is better adapted to describe electron ionization and attachment, pure DFT Kohn-Sham orbitals are preprepared to describe 
excitation energies in the sense that the occupied and unoccupied orbitals see the same potential, thus minimizing
orbital relaxation effects.  Inspection of the sizes and signs of the integrals in Eq.~(\ref{eq:results.1}) indicates
that we should expect,
\begin{equation}
  \hbar \omega_{i \rightarrow a}^T \leq \Delta \epsilon_{i \rightarrow a} \leq \hbar \omega_{i \rightarrow a}^S \, .
  \label{eq:results.3}
\end{equation}
These is confirmed in Table~\ref{tab:co-excitation-energy} where the $1^3\Pi$ and $1^1\Pi$ excitations
are, respectively, the triplet and singlet states corresponding to the HOMO $\rightarrow$  LUMO transition.

\begin{figure}[h!]
\small
\begin{center}
\includegraphics[angle=0,width=0.6\textwidth]{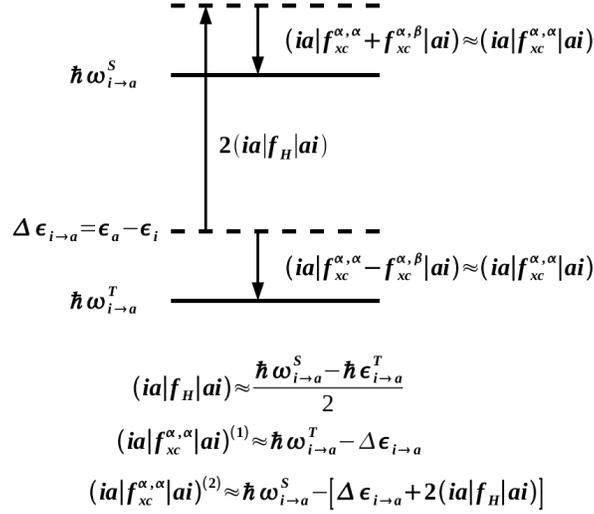}
\end{center}
\caption{Estimation of integrals within the TOTEM model.  \label{fig:TOTEM}
}
\end{figure}
Assuming that $f_{xc}^{\alpha,\alpha}$ dominates over $f_{xc}^{\alpha,\beta}$, we may even go a bit further
to estimate $(ia \vert f_H \vert ai)$ and $(ia \vert f_{xc}^{\alpha,\alpha} \vert ai)$ (Fig.~\ref{fig:TOTEM}.)
The calculations are show in Table~\ref{tab:TOTEM}.  Comparison of $(ia \vert f_{xc}^{\alpha,\alpha} \vert ai)^{(1)}$
and $(ia \vert f_{xc}^{\alpha,\alpha} \vert ai)^{(2)}$ provides an indication of the quality of the approximation
of neglecting the $(ia \vert f_{xc}^{\alpha,\beta} \vert ai)$ integral which in this case appears to be excellent.
The $(ia \vert f_H \vert ai)$ integrals calculated with the two programs are reasonably close.  Interestingly
the $(ia \vert f_{xc}^{\alpha,\alpha} \vert ai)$ disagree by about 0.4 eV which, though small, is not negligible.

\begin{table}[h!]
\small
  \caption{Estimations of integrals (in eV) within the TOTEM. \label{tab:TOTEM}}
  \begin{center}
  \begin{tabular}{ccc}
  \hline \hline 
   Program & {\sc deMon2k} & {\sc BigDFT} \\
  \hline
  \multicolumn{3}{c}{Input Data} \\
   $1^1\Pi$                       & 8.42    &  8.94      \\
   $\Delta \epsilon_{i \rightarrow a}$ & 6.87   & 6.90       \\
   $1^3\Pi$                       & 6.05    &  6.47      \\
  \multicolumn{3}{c}{Derived Results} \\
   $(ia \vert f_H \vert ai)$      & 1.19    & 1.24       \\
   $(ia \vert f_{xc}^{\alpha,\alpha} \vert ai)^{(1)}$ & -0.82    & -0.43    \\
  $(ia \vert f_{xc}^{\alpha,\alpha} \vert ai)^{(2)}$ & -0.83   &  -0.44   \\
   \hline \hline
  \end{tabular}
  \end{center}
\end{table}

\begin{figure}
\begin{center}
\includegraphics[angle=0,width=0.7\textwidth]{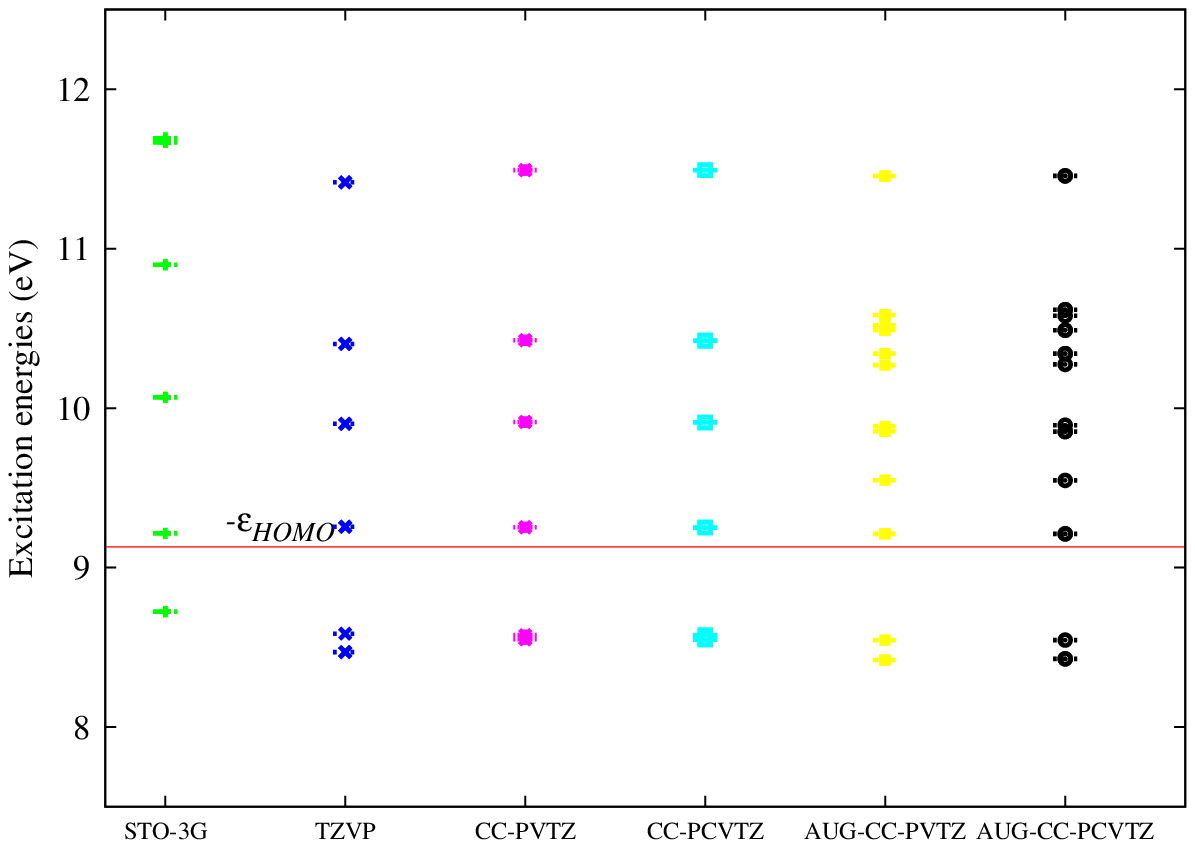}
\end{center}
\caption{ Singlet  and triplet excitation energies for CO
         calculated using {\sc deMon2k}
          \label{fig:demon-excited-states-continuum}
}
\end{figure}
Let us now examine the issue of the collapse of the continuum.  In Ref.~\cite{CJCS98}, it was shown that
the TD-DFT ionization continuum begins at $-\epsilon_{HOMO}$.  In exact Kohn-Sham DFT, this should be
the ionization potential.  However typical approximate density functionals underbind electrons and
so lead to an artificially-early on-set of the TD-DFT ionization continuum.  This is first
illustrated using the {\sc deMon2k} program and different basis sets.  Indeed 
Fig.~\ref{fig:demon-excited-states-continuum} shows that the states above $-\epsilon_{HOMO}$ tend to
collapse towards $-\epsilon_{HOMO}$ rather than converging as they should.  This is simply because
we are trying to describe a continuum which should not be there with a finite basis set.  Also seen
in the figure is a slight splitting of the $1^1\Pi$ excitation energy.  This small effect is due to
the fact that the grid used to calculate xc-integrals in {\sc deMon2k} has only roughly the symmetry
of the molecule.  

\begin{figure}[h!]
\begin{center}
\includegraphics[angle=0,width=0.7\textwidth]{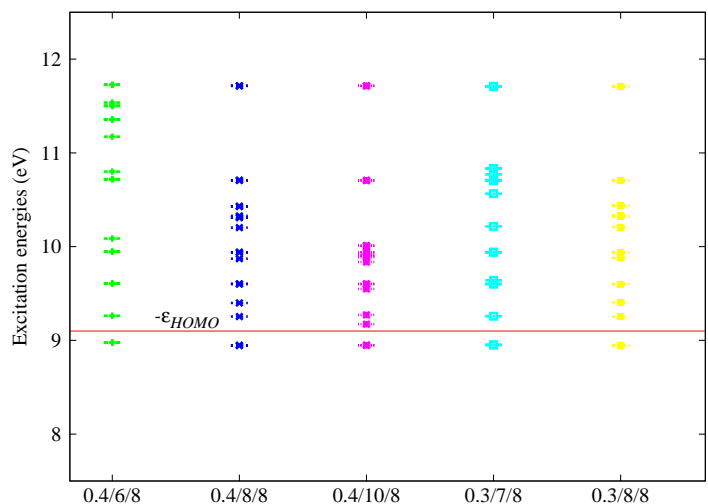}
\end{center}
\caption{Singlet  and triplet excitation energies for CO
         calculated using {\sc BigDFT} \label{fig:bigdft-excited-states-continuum}
}
\end{figure}
Now let us turn to {\sc BigDFT} calculations.  Figure~\ref{fig:bigdft-excited-states-continuum}
shows a similar collapse of the continuum as the fineness of the grid increases.  Interestingly
there is no evidence of symmetry breaking of the doubly-degenerate $1^1\Pi$ state.

\subsection{Oscillator Strengths}

Carbon monoxide is very unusual for small molecules in that absolute oscillator strengths have been
well studied \cite{CCB93} over a significant energy range and the $A^1\Pi$ ($1^1\Pi$ in 
Table~\ref{tab:co-excitation-energy}) is bright and has an accurately determined oscillator strength.
See Fig.~2 of Ref.~\cite{CCB93} (as well as other references in the same paper) for a graph of
measured absolute optical oscillator strengths against absorption energy in eV.  Table~\ref{tab:f}
reports our calculated TD-LDA/TDA oscillator strengths.  As the TDA violates the Thomas-Reiche-Kuhn (TRK) 
$f$-sum rule \cite{C95} it should only be used very cautiously to estimate oscillator strengths.  Nevertheless
the {\sc deMon2k} value of $f=0.232$ is in good agreement with the experimental value of $f=0.1762$.
As shown
in Ref.~\cite{CS00}, full TDLDA calculations with asymptotically corrected potentials give smaller oscillator
strengths (0.136 for TD-LDA/LB94 and 0.156 for TD-LDA/AC-LDA calculations\cite{CS00}.)
(Co\"{\i}ncidently our own {\sc deMon2k} 
full TD-LDA calcuations without asymptotic corrections give a degeneracyc-weighted oscillator strength
of 0.1752 (bang on the experimental value) but
an excitation energy of 8.19 eV.)
Since oscillator strengths are quite sensitive to configuration mixing with nearby
states, the fact that the {\sc BigDFT} oscillator strength is larger than the {\sc deMon2k} oscillator
strength may be due to the relatively small energy separation between the {\sc BigDFT} $A^1\Pi$ state
and the artificially-low TD-LDA ionization continuum.

\begin{table}
  \caption{Comparison of experimental $A^1\Pi$ energies (eV) and oscillator strengths with TD-LDA/TDA experimental $A^1\Pi$ 
           energies (eV) and degeneracy-weighted oscillator strengths (unitless.)\label{tab:f} }
  \begin{center}
  \begin{tabular}{cccc}
  \hline \hline
            &  {\sc deMon2k} & {\sc BigDFT} & Experiment$^a$ \\
  \hline 
  $\hbar \omega_S$ & 8.43 & 8.95 & ~8.4 \\ 
  $f$              & 0.232 & 0.853 & 0.1762 \\
  \hline \hline
  \end{tabular}
  \end{center}
$^a$ See Table VIII of Ref.~\cite{CS00}.
\end{table}   


\section{Conclusion}
\label{sec:conclude}

In this chapter we have tried to give an informative elementary review of a subject largely unfamiliar
to most theoretical chemists and physicists.  Wavelets, once an obscure ripple at the exterior of engineering
applications, grew to become a regular tsumani in engineering circles in the 1990s as the similarity to and
superiority over Fourier transform methods for multiresolution problems with arbitrary boundary conditions
became increasingly recognized.  Though the first applications of wavelet theory to solving the Schr\"odinger equation
may be traced back to the mid-1990s \cite{A99,FD93,C96}, the theory is
still not well known among quantum mechanicians.  Here we have tried to remedy this aberrant situation by trying to 
``make some {\em waves} about {\em wavelets} for {\em wave} functions.''

In particular we have reviewed the theory behind the wavelet code {\sc BigDFT} for ground-state DFT and our
recent implementation of wavelet-based TD-DFT in {\sc BigDFT}.  Rapid progress is being made towards making
{\sc BigDFT} a high performance computing  order-$N$ code for applications to large systems.  Right now
applications to 400 or 500 atoms are routine for ground-state calculations with {\sc BigDFT}.  Our implementation
of TD-DFT in {\sc BigDFT} is by comparison only a rudementary beginning, but it shows that the basic method
is viable and we are confident that there are no insurmountable obstacles to making high performance computing order-$N$
wavelet-based TD-DFT code for large systems.

\section*{Acknowledgments}

B.\ N.\ would like to acknowledge a scholarship from the {\em Foundation Nanoscience}.
This work has been carried out in the context of the French Rh\^one-Alpes
{\em R\'eseau th\'ematique de recherche avanc\'ee (RTRA): Nanosciences aux limites de la
nano\'electronique} and the Rh\^one-Alpes Associated Node of the European Theoretical
Spectroscopy Facility (ETSF).


\section*{Appendices}
\appendix
\section{List of Abbreviations}


For the readers convenience we give a list of the abbreviations used in this chapter in
alphabetical order:
\begin{description}
\item[AA] Adiabatic approximation.
\item[APW] Augmented plane wave.
\item[CC] Coupled cluster.
\item[CI] Configuration interaction.
\item[DFT] Density-functional theory.
\item[DM] Density matrix.
\item[FD] Finite difference.
\item[FE] Finite element.
\item[FEM] Finite element method.
\item[GTH-HGH] Goedecker-Teter-Hutter/Hartwigsen-Goedecker-Hutter.
\item[GTO] Gaussian-type orbitals.
\item[H] Hartree.
\item[HF] Hartree-Fock.
\item[HOMO] Highest-occupied molecular orbital.
\item[ISF] Interpolating scaling function.
\item[LCAO] Linear combination of atomic orbitals.
\item[LDA] Local density approximation.
\item[LMTO] Linear muffin tin orbital.
\item[LR] Linear-response.
\item[LR-TD-DFT] Linear-response time-dependent density-functional theory.
\item[LR-TD-LDA] Linear-response time-dependent local density approximation.
\item[LUMO] Lowest-unoccupied molecular orbital.
\item[NS] Non-standard.
\item[MBPT] Many-body perturbation theory.
\item[MRA] Multiresolution analysis.
\item[S] Singlet.
\item[SOS] Sum-over-states.
\item[STO] Slater-type orbital
\item[T] Triplet.
\item[TD] Time-dependent.
\item[TDA] Tamm-Dancoff approximation.
\item[TD-DFT] Time-dependent density-functional theory.
\item[TD-LDA] Time-dependent local density approximation.
\item[TRK] Thomas-Reiche-Kuhn.
\item[xc] Exchange-correlation.
\end{description}

\bibliographystyle{myaip}
\bibliography{refs}
\label{lastpage-01}
\end{document}